\documentclass[12pt,letter]{article}
\pdfoutput=1
\usepackage{graphicx, epsfig, color,cite}
\usepackage{amsmath}
\usepackage{amssymb}
\usepackage{float}
\usepackage{subfig}
\usepackage{hyperref}

\def\eslt{\not\!\!\!{E_T}}
\def\to{\rightarrow}

\def\bi{\begin{itemize}}
\def\ei{\end{itemize}}

\def\tg{\tilde g}

\def\tw{\widetilde\chi^{\pm}}
\def\tz{\widetilde\chi^0}
\def\alt{\lesssim}
\def\agt{\gtrsim}
\def\be{\begin{equation}}  
\def\ee{\end{equation}}  
\def\bea{\begin{eqnarray}}  
\def\eea{\end{eqnarray}}

\begin{document}
\begin{titlepage}
\begin{flushright}
OU-HEP-231104
\end{flushright}

\vspace{0.5cm}
\begin{center}
  {\Large \bf Winos from natural SUSY at the high luminosity LHC}\\
\vspace{1.2cm} \renewcommand{\thefootnote}{\fnsymbol{footnote}}
{\large Howard Baer$^{1}$\footnote[1]{Email: baer@ou.edu },
Vernon Barger$^2$\footnote[2]{Email: barger@pheno.wisc.edu},
Xerxes Tata$^3$\footnote[3]{Email: tata@phys.hawaii.edu} and
Kairui Zhang$^2$\footnote[5]{Email: kzhang89@wisc.edu}
}\\ 
\vspace{1.2cm} \renewcommand{\thefootnote}{\arabic{footnote}}
{\it 
$^1$Homer L. Dodge Department of Physics and Astronomy,
University of Oklahoma, Norman, OK 73019, USA \\[3pt]
}
{\it 
$^2$Department of Physics,
University of Wisconsin, Madison, WI 53706 USA \\[3pt]
}
{\it 
$^3$Department of Physics and Astronomy,
University of Hawaii, Honolulu, HI 53706 USA \\[3pt]
}

\end{center}

\vspace{0.5cm}
\begin{abstract}
\noindent
In natural supersymmetric models defined by no worse than a part in
thirty electroweak fine-tuning, winos and binos are generically expected
to be much heavier than higgsinos. Moreover, the splitting between the
higgsinos is expected to be small, so that the visible decay products of
the heavier higgsinos are soft, rendering the higgsinos quasi-invisible
at the LHC. Within the natural SUSY framework, heavy electroweak gauginos
decay to $W$, $Z$ or $h$ bosons plus higgsinos in the ratio $\sim
2:1:1$, respectively. This is in sharp contrast to models with a
bino-like lightest superpartner and very heavy higgsinos, where the charged
(neutral) wino essentially always decays to a $W$ ($h$) boson and an
invisible bino.  Wino pair production at the LHC, in natural SUSY, thus
leads to $VV$, $Vh$ and $hh+\eslt$ final states ($V=W,Z$) where, for TeV
scale winos, the vector bosons and $h$ daughters are considerably
boosted.  We identify eight different channels arising from the leptonic
and hadronic decays of the vector bosons and the decay $h\to b\bar{b}$,
each of which offers an avenue for wino discovery at the high luminosity
LHC (HL-LHC). By combining the signal in all eight channels we find,
assuming $\sqrt{s}=14$ TeV and an integrated luminosity of 3000
fb$^{-1}$, that the discovery reach for winos extends to $m(wino)\sim
1.1$~TeV, while the 95\% CL exclusion range extends to a wino mass of
almost 1.4~TeV. We also identify ``higgsino specific channels'' which
could serve to provide $3\sigma$ evidence that winos lighter than
1.2~TeV decay to light higgsinos rather than to a bino-like LSP, should
a wino signal appear at the HL-LHC.

\end{abstract}
\end{titlepage}

\section{Introduction}
\label{sec:intro}
The search for supersymmetric partners of Standard Model (SM) particles
at high energy colliders has been on the cutting edge of high energy
physics ever since it was realized that supersymmetry could stabilize
the weak scale provided that the superpartner masses were not very far
above the TeV scale
\cite{Witten:1981nf,Dimopoulos:1981zb,Sakai:1981gr,Kaul:1981hi}. Indeed
the discovery \cite{ATLAS:2012yve,CMS:2012qbp} of the seemingly SM-like
Higgs boson makes it even more urgent to discover why radiative
corrections do not drive its mass to the scale of the most massive
particles that couple to the particles of the SM.  Even ignoring gravity
on the grounds that we may not know how to incorporate it into a quantum
framework, there are a number of reasons to suppose that there are new
particles (that couple to the Higgs sector) with masses between the weak
scale and the Planck scale. Though it remains a speculation at the
present time, arguably the most compelling theoretical reason for new
particles is the Grand Unification of the electroweak and strong
interactions into a single gauge interaction at the scale $M_{\rm
  GUT}$. Non-zero neutrino masses may also have their origin in heavy
(SM singlet) particles if these acquire their masses via the so-called
see-saw mechanism as opposed to tiny dimensionless Yukawa couplings:
implications of this for the hierarchy problem are discussed in
Ref.\cite{Vissani:1997ys}.  There could also be new particles at a scale
associated with the origin of flavour. Regardless of what the new
physics is, radiative corrections would generically make the Higgs boson
squared mass {\em quadratically sensitive} to this new scale, except in
a supersymmetric theory with the SUSY breaking scale well below the
scale associated with the new physics: in this case, the Higgs boson
mass squared would only be logarithmically sensitive to the UV scale,
but quadratically sensitive to the scale of SUSY breaking.

These considerations had led to much hope that superpartners would show
up in direct searches at the LHC. As is well known, this has not
happened and the non-observation of an excess of events in various
channels have led only to lower limits of $\sim 2.3$~TeV on the
masses of strongly interacting gluinos decaying to third generation
quarks, of 1.4-1.8~TeV on squarks (assuming an approximate degeneracy
among squark flavours) and about 1.3~TeV on the top squark
\cite{ATLAS:2021twp,ATLAS:2022ihe,CMS:2019ybf,CMS:2021beq}.
There are also lower bounds of several hundred GeV to just over a TeV on
the masses of electroweakly interacting sleptons and the winos
\cite{CMS:2022sfi,CMS:2021few,CMS:2021cox,CMS:2023qhl,ATLAS:2020pgy,ATLAS:2021yqv,ATLAS:2022hbt,ATLAS:2022zwa}. It
should be noted though that these limits are mostly obtained in
simplified models, assuming $R$-parity conservation, specific decay
modes of the parent sparticle,  and a large mass gap
between the particle being searched for and the lightest supersymmetric
particle (LSP) often assumed to be the lightest
neutralino.\footnote{There are also stringent limits on sparticle masses
  in $R$-parity violating models, but these will not concern us in this
  paper.}

The absence of signals at the LHC has led some authors to suggest
that the supersymmetry is unable to explain the value
of the Higgs boson mass without
resorting to some degree of fine-tuning, typically stated to be at a
parts per mille level. These authors evaluate the sensitivity of
$m_Z^2$, which serves the electroweak scale and is calculable in terms
of model parameters, to the {\em independent} parameters, $a_i$, of the
model: $\Delta = max_i|\frac{a_i}{M_Z^2}\frac {\partial M_Z^2}{\partial
  a_i}|$ \cite{Ellis:1986yg,Barbieri:1987fn,Dimopoulos:1995mi,Feng:2013pwa}.
Typically, most of the parameters in any model $a_i$ have to do
with soft supersymmetry breaking (SSB),
with the superpotential higgsino mass, $\mu$, often the
sole dimensionful supersymmetric parameter.  In the absence of an
understanding of how superpartners acquire their masses, it is
not possible to know how the SSB parameters are correlated in the underlying
theory. It has, however, been pointed out that ignoring correlations
among the parameters can lead one to overestimate the degree of
fine-tuning (by up to a factor $10^3$\cite{Baer:2023cvi}),
and prematurely cause us to discard perfectly viable
models\cite{Baer:2013gva,Mustafayev:2014lqa}. For this reason, we follow
a different path to assess the fine-tuning in the minimal supersymmetric
Standard Model (MSSM).

Again, we take the experimental value of the $Z$-boson mass to represent the
magnitude of weak scale, but this time express it in terms of Lagrangian
parameters determined at the weak scale via the 
minimization of the  potential in the Higgs sector as
\be m_Z^2/2
=\frac{m_{H_d}^2+\Sigma_d^d-(m_{H_u}^2+\Sigma_u^u
  )\tan^2\beta}{\tan^2\beta -1}-\mu^2.
\label{eq:mzs}
\ee 
Here, $m_{H_u}^2$ and $m_{H_d}^2$ are the Higgs soft breaking mass
terms, $\mu$ is the (SUSY preserving) superpotential higgsino mass
parameter, and the $\Sigma_d^d$ and $\Sigma_u^u$ terms include an
assortment of loop corrections that are typically most sensitive to third
generation sfermion and gaugino masses (see Appendices of
Ref. \cite{Baer:2012cf} and \cite{Baer:2021tta} and also see
\cite{Dedes:2002dy} for leading two-loop corrections).  We then require
that none of the individual terms on the right-hand-side are much larger
than $m_Z^2/2$, {\it i.e.} there are no large cancellations necessary
between the supersymmetric term $\mu^2$ and the SSB terms (or for that
matter between the SSB terms in various sectors) that presumably have
very different physics origins \cite{Bae:2019dgg}. With this in mind, we
use the electroweak fine-tuning measure
measure\cite{Baer:2012cf,Baer:2012up}
\be \Delta_{EW}\equiv |{\rm
  maximal\ term\ on\ the\ right \  hand \ side \ of\ Eq.~(\ref{eq:mzs})}|/(m_Z^2/2)\;.
\ee 

It has been argued that (modulo some technical caveats), $\Delta_{EW}
\le\Delta$, with the inequality being saturated only for specific
correlations between the parameters\cite{Mustafayev:2014lqa}, and
further, that $\Delta_{EW}$ measures the minimum fine-tuning for a given
sparticle spectrum.  This makes $\Delta_{EW}$ a very conservative
estimate of fine-tuning and precludes the possibility of discarding a
model even in the presence of correlations among the parameters.  We
adopt $\Delta_{EW}\alt 30$ as our criterion for naturalness.

We note here that it has been suggested that notions of stringy
naturalness\cite{Baer:2019cae} applied to the landscape of string theory
vacua, together with the anthropic requirement that a diversity of
nuclei form -- this anthropic requirement requires that the value of
$m_Z$ be no larger than a factor $\sim 4$ of its observed value
\cite{Agrawal:1998xa} -- lead to SUSY models with values of
$\Delta_{EW}\alt 30$ and heavy superpartners (other than light
higgsinos): see Ref.\cite{Baer:2020kwz} for a review, and for references
to the original literature. The reader who does not subscribe to these
landscape considerations may view $\Delta_{EW} \alt 30$ as in between 10
(accidental cancellations of an order of magnitude are known, {\it e.g.}
the decay rate of orthopositronium includes a factor of $\pi^2-9$)
and 100 (presumably too large to be attributed to an accidental
cancellation). The reader who does not wish to entertain any notions of
naturalness may disregard the discussion of the last two paragraphs, and
view our analysis as a search for winos in models with light higgsinos,
{\it i.e} where $|\mu| \ll M_{1,2}$, where $M_{1,2}$ are the bino and
wino mass parameters at the weak scale.

Requiring $\Delta_{EW}<30$ (this ensures small $\mu$) immediately
implies that each of the contributions in Eq.~(\ref{eq:mzs}) is no
bigger than a factor of a few relative to $m_Z$. Specifically:
\begin{itemize}

\item The $\mu$ parameter has a magnitude
smaller than $\sim 350$~GeV, so that the higgsinos are expected to be in
the 110-350~GeV range, with the lightest neutral higgsino being the LSP.

\item The finite radiative corrections $\Sigma_u^u$ have the same
  upper bound, which requires the top squarks to be bounded above by
  $\sim 3$~TeV and the gluino by $\sim 6-9$~TeV\cite{Baer:2015rja},
  so these can all be
  well beyond the discovery reach of even the high luminosity LHC
  (HL-LHC), about 2.8~TeV for the gluino\cite{Baer:2016wkz} and
  1.3-1.7~TeV\cite{CidVidal:2018eel,Baer:2023uwo} for the stop,
  depending on how the stop is assumed to decay.

\item Wino masses enter Eq.~(\ref{eq:mzs}) only via loop corrections and
  can be in the 4-5~TeV range without endangering naturalness
  \cite{Baer:2015rja}, though in models with gaugino mass unification
  their mass is somewhat more tightly constrained by the naturalness limit on
  $m_{\tg}$ noted above. In any event, the  magnitude of the
  wino mass parameter is expected to be much larger than
  $|\mu|$. As a result, the lighter neutralinos $\tz_{1,2}$ and the
  lighter chargino $\tw_1$ are expected to be higgsino-like while the
  heavier neutralinos $\tz_{3,4}$ and the heavy chargino $\tw_2$ are
  expected to be gaugino-like. The upper limits on the wino and bino
  masses are also important in that they severely limit the splitting
  between the higgsinos -- a small higgsino mass gap results in very
  soft visible debris from the decay of the heavier higgsinos to the
  LSP. As a result, higgsino pair production, in spite of its large
  production cross section at the LHC, is swamped by SM
  backgrounds\cite{Baer:2011ec}.  It has been suggested that higgsino
  production in association with a hard QCD jet that leads to monojet
  plus $\eslt$ events with soft dilepton events from the decay
  $\tz_2\to\tz_1\ell\bar{\ell}$ leads to a viable signal
  \cite{Han:2014kaa,Baer:2014kya,Han:2015lma}. Both ATLAS and CMS have
  explored this channel but up to now have excluded a sizeable portion
  of the $\mu$-$\Delta m$ plane ($\Delta m=m_{\tz_2}-m_{\tz_1}$) allowed
  by naturalness considerations\cite{ATLAS:2019lng,CMS:2021edw}.

\item First and second generation squarks and sleptons with restrictions
  on intra-generation splittings are only weakly constrained by
  naturalness. They  can range up to
  ${\cal O}(40)$~TeV without jeapordizing naturalness, greatly
  ameliorating the SUSY flavour problem \cite{Baer:2013jla,Baer:2019zfl}.

\end{itemize}

As noted above, in the context of natural SUSY considerable effort has
been expended on the search for higgsinos because higgsino masses are
bounded above.  There have also been many LHC searches for electroweak
gauginos though these have mostly been carried out within the context
of simplified models with a bino-like LSP. These analyses typically
assume that the wino-like chargino decays via $\tw_1 \to W\tz_1$ and
the heavier neutralino decays via $\tz_2\to Z\tz_1$ or $\tz_2\to
h\tz_1$.  They also assume that higgsinos are too heavy to be produced
at the LHC.\footnote{Strictly speaking if $\tz_2$ is wino-like and
  $\tz_1$ is bino-like, the decay $\tz_2\to h\tz_1$ would dominate
  $\tz_2\to \tz_1 Z$ because (as explained in Sec.~\ref{sec:higgsino}),
  the latter
can occur only via the suppressed higgsino components of {\em both}
$\tz_1$ and $\tz_2$, while the decay to the Higgs boson requires just
a single mixing angle suppression. The importance of the $W(\to
\ell\nu)h(\to b\bar{b})$ signal from wino production at the LHC was
first pointed out in Ref.\cite{Baer:2012ts}.} The signals with the
lowest backgrounds come from the leptonic decays of the vector bosons
and lead to trilepton events +$\eslt$ events with hadronic activity
only from QCD radiation, and lead to a lower limit on the wino mass
$\sim 650$~GeV assuming a light LSP\cite{ATLAS:2021moa}.
In the model where the chargino
decays via $\tw_1 \to W(\to\ell)\tz_1$ while the neutralino decays via
$\tz_2\to h(\to b\bar b) \tz_1$, the wino limit extends to about
750-800~GeV for a light LSP\cite{ATLAS:2018qmw}.
Remarkably, the strongest limits on the
wino mass arise from the {\em hadronic} decays of the $W$, $Z$ and
$h$ bosons, and excludes winos lighter than $\sim 1$~TeV for a bino
LSP as heavy as 300~GeV \cite{CMS:2023qhl,ATLAS:2021yqv}.

In this paper, we examine the reach of the HL-LHC for winos in the
context of natural SUSY.\footnote{For a survey of chargino and neutralino
signals in the $\mu-M_2$ plane assuming a decoupled bino, see
Ref.~\cite{Carpenter:2023agq}.}
These winos will for the most part decay
into the lighter higgsinos plus a $W$, $Z$ or $h$, with branching
fractions in the ratio 2:1:1.  Our focus will be on the wino states
$\tw_2$ and $\tz_4$ because the bino-like $\tz_3$ couples to gauge
bosons (we assume that squarks are very heavy) only via
mixing.\footnote{For definiteness, we assume gaugino mass unification
  which makes the bino lighter than the neutral wino in our
  calculations. In models where the bino is heavier than the wino, our
  results on wino signals will be qualitatively unaltered (keep in mind
  that wino decays to the bino-like state are suppressed by mixing
  angles) if we remember to interchange $\tz_3 \leftrightarrow \tz_4$. }
As already noted in Ref.\cite{Baer:2018hpb}, wino pair production thus
leads to $VV$, $Vh$ and $hh$ $+\eslt$ ($V=W,Z$) channels via which to
search for SUSY at high energy colliders. Motivated by the ATLAS and CMS
analyses, we examine signals from both leptonic as well as hadronic
decays of the daughter $W$ and $Z$ boson daughters of the winos.

The analysis methods developed in the early days of supersymmetry to
search for winos decaying to binos (assuming decoupled higgsinos) must be
adapted for the search for heavy winos of natural SUSY as long as
the visible decay products of the daughter higgsinos are assumed to be
too soft for detection. We, of course, need to keep track of the
branching fractions of the charged and neutral winos to decay via the
$W, Z$ or $h$ channels. The cleanest channels -- which come from the
leptonic decays of the bosons and yield events with up to four hard
leptons plus $\eslt$ -- were examined in Ref.\cite{Baer:2013xua} a decade
ago. It should be straightforward for the ATLAS and CMS collaborations to
incorporate the branching fractions for wino decays predicted by natural
SUSY models and repeat their analyses to obtain wino mass bounds within
this presumably more realistic/plausible framework. In the absence of a SUSY
signal, this may not seem essential. The situation will be very
different if a signal appears in Run 3 or in the HL-LHC run in the
future because the model {\em predicts} relative rates for the signals in
various leptonic channels as well as in mixed hadron-lepton channels and
purely hadronic channels, with and without $b$-tags.

The remainder of this paper is organized as follows. In the next section
we introduce the non-universal Higgs mass model that we use for the
analysis of the wino signal in natural SUSY models at the HL-LHC. In
Sec.~\ref{sec:prod}, we discuss the production cross sections for
electroweakino (EWino) production at a $pp$ collider with $\sqrt{s}=14$~TeV.
In Sec.~\ref{sec:BF} we map out the decay patterns of EWinos
in models with small values of $|\mu|$, one of the key characteristics
of natural SUSY models. In Sec.~\ref{sec:channels} we delineate the
various channels and detail the analysis cuts that we use to optimize
the search for winos. In Sec.~\ref{sec:reach}, we present our results
for the HL-LHC discovery reach for winos in natural SUSY models. In
Sec.~\ref{sec:higgsino}, we discuss whether an examination of the HL-LHC
signal from winos by itself can provide evidence for the existence of a
higgsino-like LSP. We end in Sec.~\ref{sec:conclude} with a summary of
our results and some general remarks.

\section{A natural SUSY model line}
\label{sec:model}

For our phenomenological analysis of wino signals, we use the two extra
parameter non-universal Higgs model (NUHM2)
\cite{Matalliotakis:1994ft,Ellis:2002wv,Nath:1997qm} specified by parameters,
$$ m_0,m_{1/2},A_0,\tan\beta,\mu, m_A .$$ The universal SUSY breaking
matter scalar mass parameter, $m_0$, the universal SUSY breaking gaugino
mass parameter, $m_{1/2}$, and the universal SUSY breaking trilinear
scalar coupling, $A_0$ are all specified at the GUT scale while the
remaining three parameters are specified at the weak scale.\footnote{The
  NUHM2 framework allows for independent soft SUSY breaking Higgs mass
  parameters $m_{H_u}^2$ and $m_{H_d}^2$ at the GUT scale. These have
  been traded in for the weak scale parameters $\mu$ and $m_A$. The
  assumed universality of matter scalar mass parameters ameliorates
  unwanted flavour-changing effects.}  This form of the NUHM2 parameter
space is very convenient for studies of natural SUSY because $|\mu|$ can
be chosen to be in the 100-350~GeV range as required. For our wino
analysis we adopt the model-line,
\begin{equation}
m_0= 5\ {\rm TeV}, m_{1/2}, A_0=-1.6m_0, \tan\beta=10, \mu=250 \ {\rm
  GeV}\ {\rm and}\ \ m_A=2~{\rm TeV}
\label{eqn:mline}
\end{equation}
which ensures modest values of $\Delta_{EW}$ along with a value of the
light Higgs boson mass $m_h=125 \pm 2$~GeV. We have, therefore, dubbed
this the $m_h^{125}(nat)$ scenario. By changing $m_{1/2}$ we can vary
the mass of the wino.

We use the computer code ISAJET \cite{Paige:2003mg} which includes
Isasugra to obtain the sparticle spectrum and weak scale couplings
relevant for phenomenology. A sample spectrum for $m_{1/2}=1.2$~TeV
which yields $\Delta_{EW}=22$ is listed in Table 1 of
Ref.\cite{Baer:2022qqr}.

We should mention that although we are using the NUHM2 framework with
unified gaugino mass parameters, this aspect plays very little role
in our examination of the phenomenology of winos at HL-LHC. We do
not look at gluino events, and except for relatively small values of
$m_{1/2}$ where the bino mass parameter is small enough so that mixing
with higgsinos is sizeable, the bino state plays no role in our
analysis since matter sfermions are taken to be heavy.  For all
practical purposes, $m_{1/2}$ only serves to determine the mass of the
wino. Finally, since we make no attempt to look at the soft debris
from higgsino decays, our results should only be mildly sensitive to the
choice of $\mu=250$~GeV, at least for wino mass much larger than
$|\mu|$.
For convenience, we note that the weak scale wino mass $M_2\sim 0.8 m_{1/2}$.

\section{EWino production cross sections at LHC14}
\label{sec:prod}

We begin by considering the production cross sections for pair
production of EWinos at the LHC. For our calculation of the
NLO cross sections, we use the computer program
PROSPINO\cite{Beenakker:1996ed} using the masses and mixing angles as
given by the ISAJET\cite{Paige:2003mg} Les Houches Accord files (LHA).

\begin{figure}[htb!]
\centering
    {\includegraphics[height=0.27\textheight]{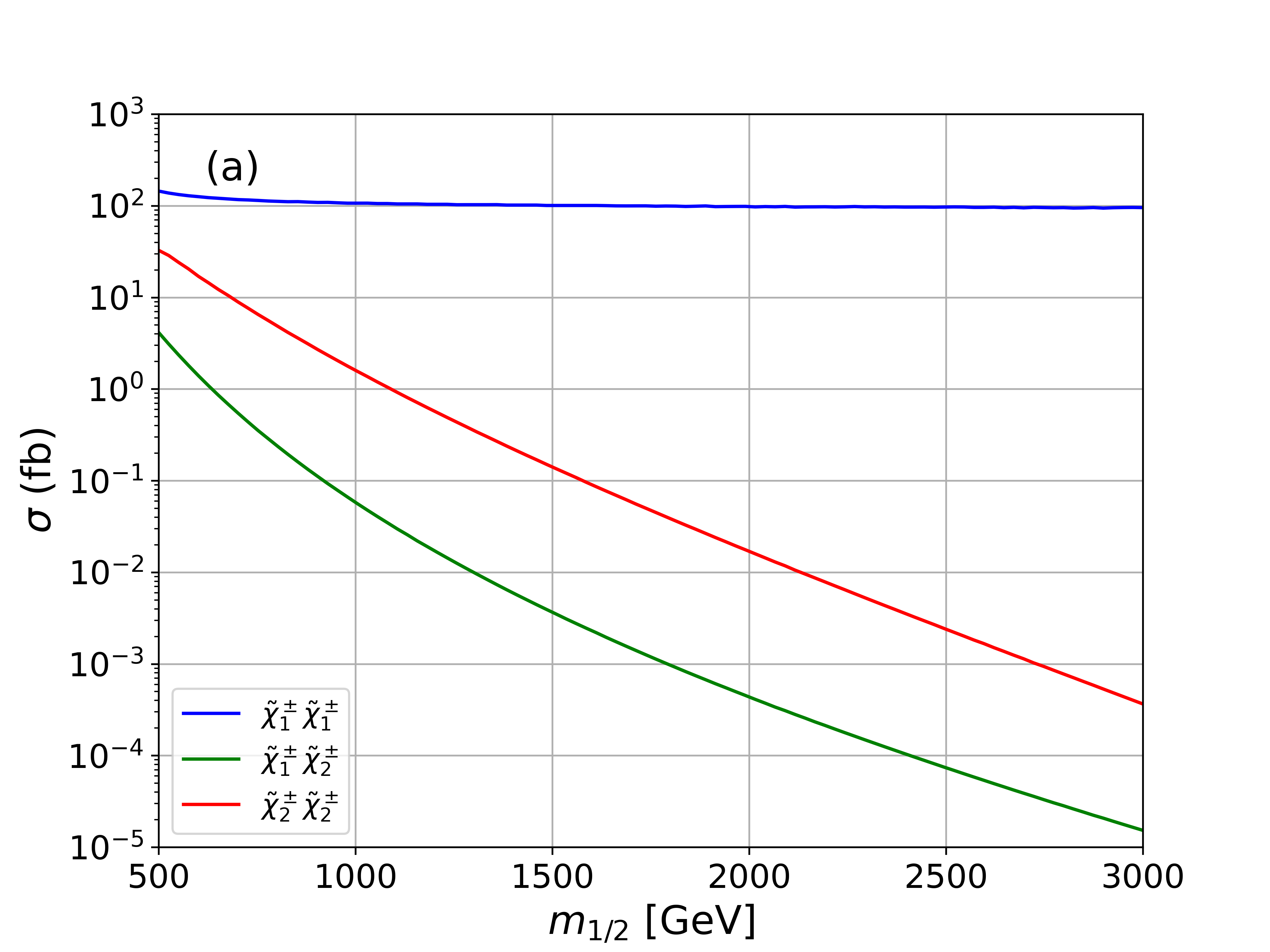}}\\
    {\includegraphics[height=0.27\textheight]{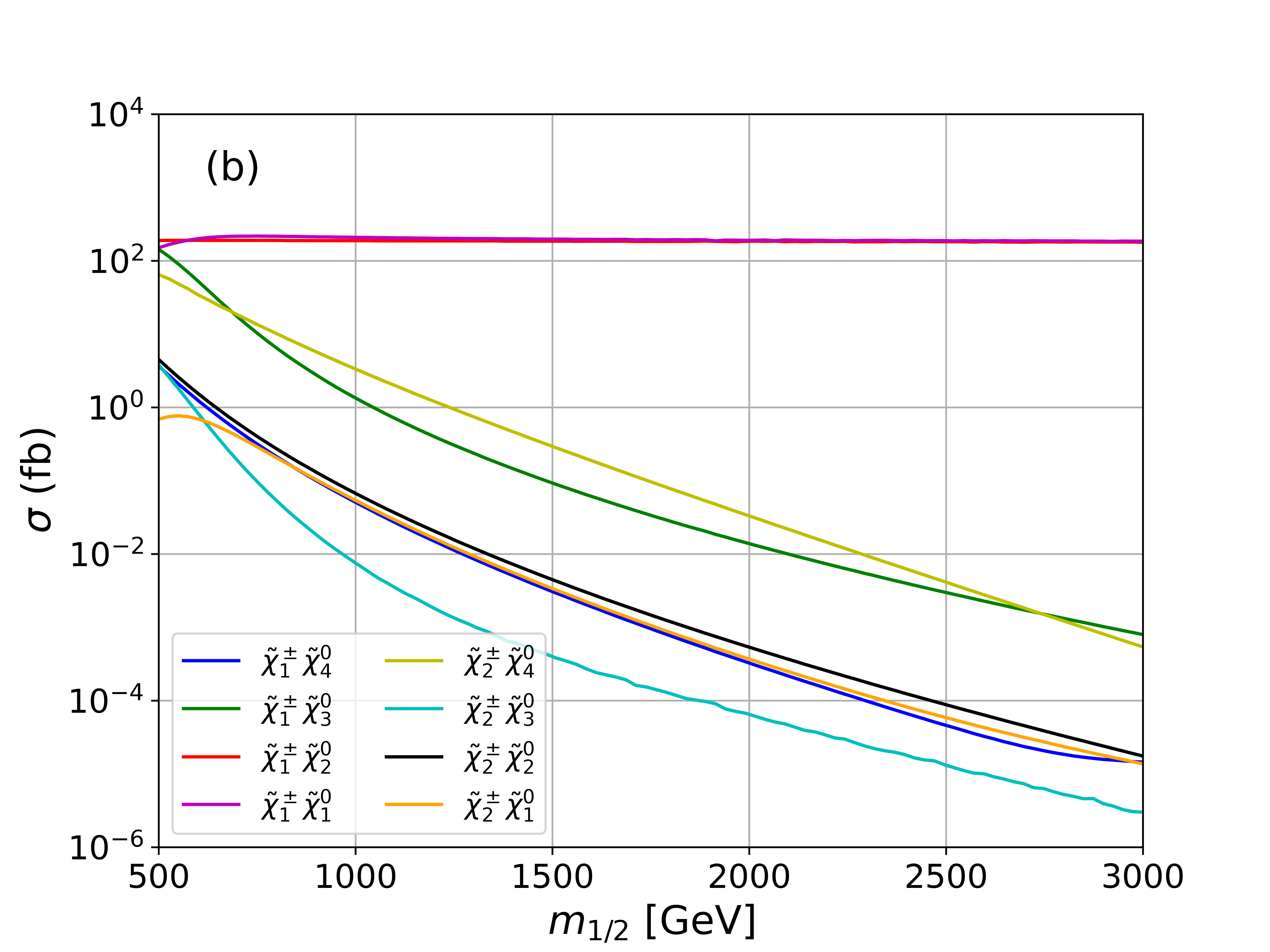}}\\
    {\includegraphics[height=0.27\textheight]{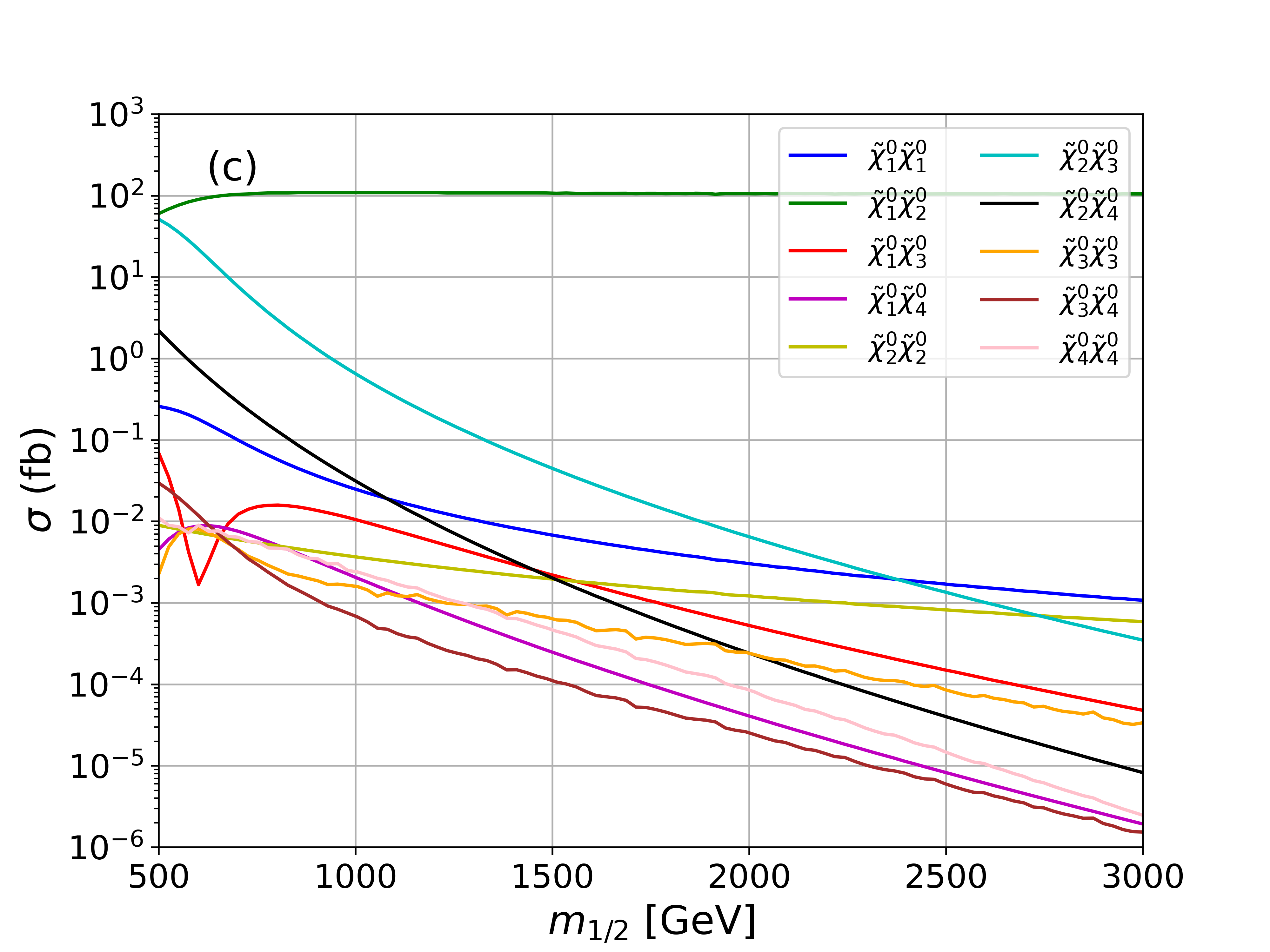}}
    \caption{NLO cross sections for EWino production at a $pp$
      collider with $\sqrt{s}=14$~TeV versus $m_{1/2}$ for the
      $m_h^{125}(nat)$ model line introduced in Eq.~(\ref{eqn:mline}) of
      the text. We show cross sections for ({\it a})~chargino pair
      production, ({\it b})~chargino-neutralino pair production, and
      ({\it c})~neutralino pair production. 
        \label{fig:prod}}
\end{figure}

Our results are shown in Fig.~\ref{fig:prod} for ({\it a})~chargino
pair production, ({\it b})~associated chargino-neutralino production,
and ({\it c})~neutralino pair production. We display the cross
sections versus $m_{1/2}$ for the $m_h^{125}(nat)$ model line defined
by Eq.~(\ref{eqn:mline}). The region with $m_{1/2}\alt 1.1$~TeV is
excluded at 95\%CL by the non-observation of any excess of events in
the wino search at the LHC
\cite{CMS:2023qhl,ATLAS:2021yqv}.\footnote{There is some slop in the
lower limit on $m_{1/2}$ which has been obtained using simplified
model analyses, with different assumptions of the branching ratios for
wino decays than in the model adopted in this paper.}

Higgsino pair production is the dominant EWino production cross section
in all the frames with a value at the hundreds of fb level.
Since the lighter EWinos $\tw_1, \tz_1$ and $\tz_2$ are
dominantly higgsino-like with a mass close to $|\mu|$ over essentially
the entire range of the plot, the cross section shows little variation
with $m_{1/2}$. Note also that pair production of identical
neutralinos is dynamically suppressed. As noted in
Sec.~\ref{sec:intro}, these higgsino pair production processes are not
directly of interest to us in this paper because the visible decay
products from higgsinos are quite soft causing the higgsino pair
signal to be difficult to extract at the LHC.

The next highest cross sections are for wino pair production processes
$\tw_2\tw_2$ and $\tw_2\tz_4$ in frames ({\it a}) and ({\it b}),
respectively. Wino pair production occurs via the large $SU(2)$ gauge
coupling and is essentially unsuppressed by mixing angles in natural
SUSY models. These are at the 1~fb level for $m_{1/2} = 1$~TeV, and, of
course, fall with increasing values of $m_{1/2}$. However, even for
$m_{1/2}$ as large as 1.8~TeV, we may expect about a hundred wino pair
events in a sample of 3000~fb$^{-1}$.  Neutral wino pairs cannot couple
to the $Z$ boson because of $SU(2)$ symmetry and so $\tz_4\tz_4$
production in frame ({\it c}) is strongly suppressed. Wino-bino
production processes, $\tz_3\tz_4$ and $\tz_3\tw_4$ are also suppressed
for reasons already mentioned in Sec.~\ref{sec:intro}.

Finally, we turn to gaugino-higgsino pair production. Since gauge bosons
couple only to higgsino pairs or gaugino pairs, gaugino-higgsino pair
production from $q\bar{q}$ collisions via virtual $W$ or $Z$
exchange in the $s$-channel is suppressed by the gaugino-higgsino
mixing. As a result, in the model with unified gaugino masses where $M_2
\simeq 2M_1$, wino-higgsino mixing is smaller than bino-higgsino
mixing. Thus the various wino-higgsino processes in the figure are
suppressed relative to the corresponding bino-higgsino processes, both
by kinematic ($M_2 > M_1$) as well as dynamical reasons ({\it e.g}
$\tw_1\tz_3$ vs. $\tw_1\tz_4$ or $\tz_2\tz_3$ vs. $\tz_2\tz_4$). Indeed,
the cross sections for some of the bino-higgsino processes, such as
$\tz_3\tw_1$ production in frame ({\it b}) or $\tz_3\tz_2$ in frame
({\it c}) are comparable in magnitude to the cross section for wino pair
($\tw_2\tw_2$ production in frame ({\it a}) or $\tw_2\tz_4$ production
in frame ({\it b})) whose LHC signatures are the subject of this paper.

\section{Electroweak gaugino branching fractions in natural SUSY}
\label{sec:BF}

\begin{figure}[htb!]
    \centering
    {\includegraphics[height=0.27\textheight]{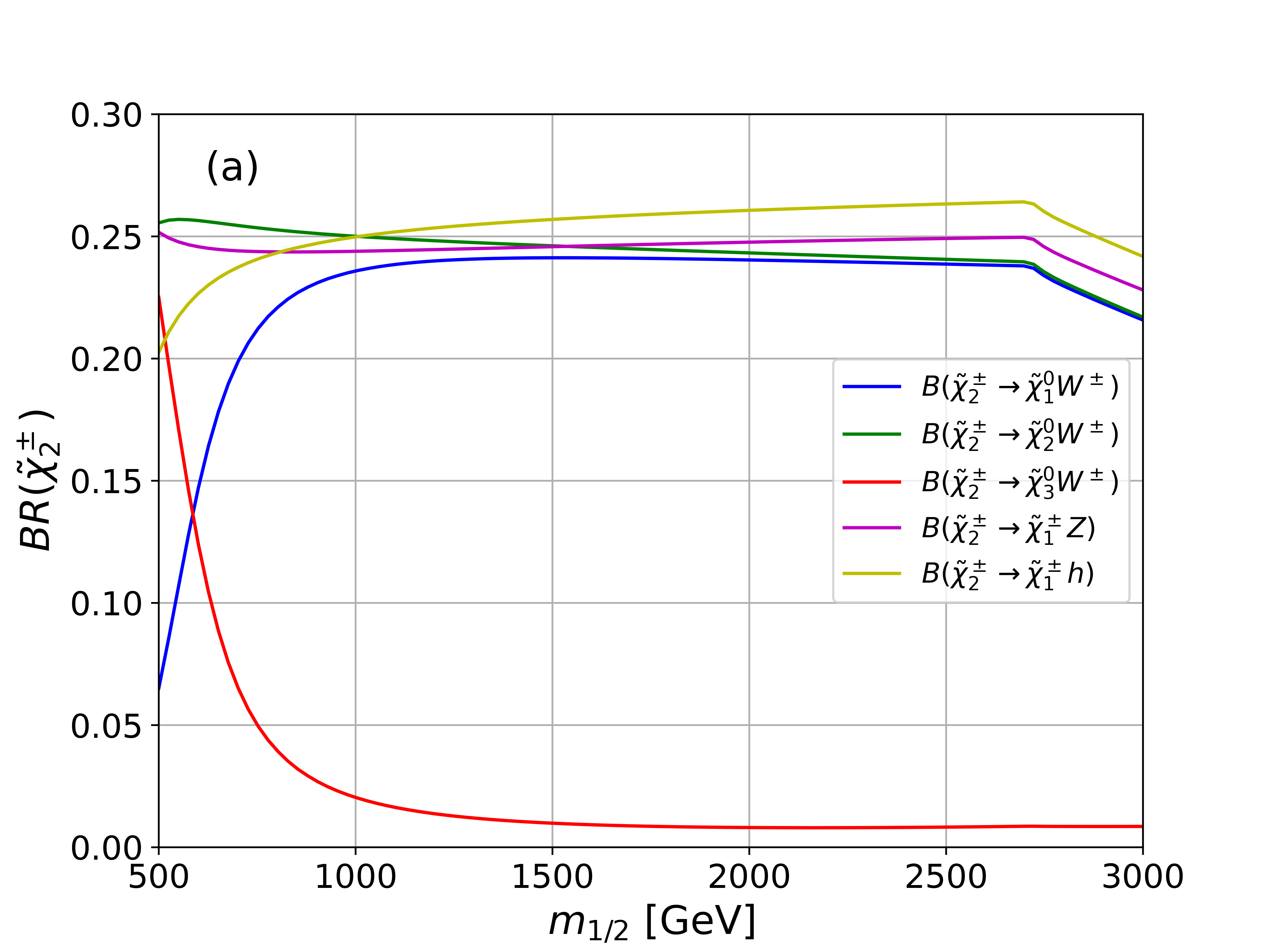}}\\
    {\includegraphics[height=0.27\textheight]{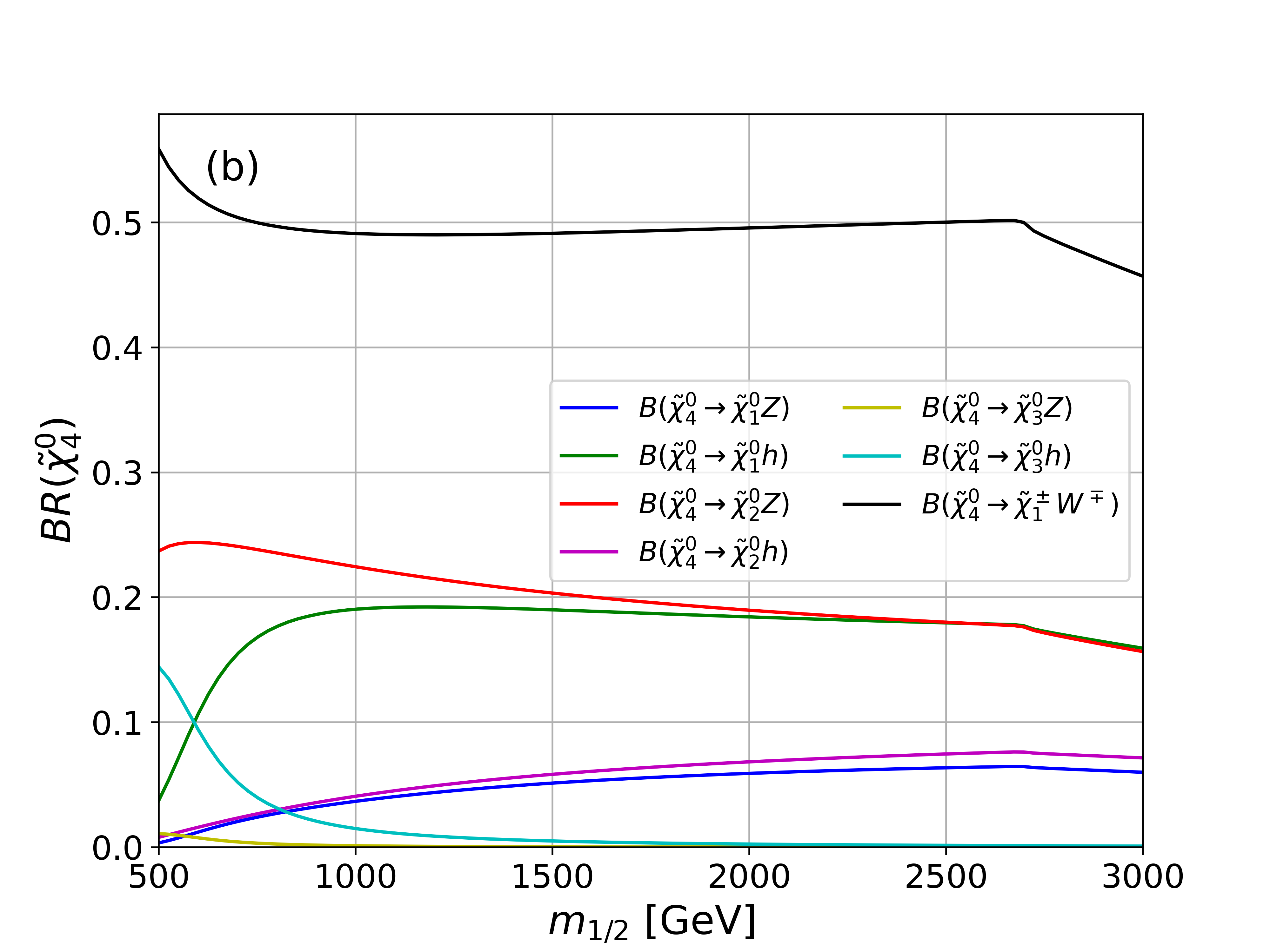}}\\
    {\includegraphics[height=0.27\textheight]{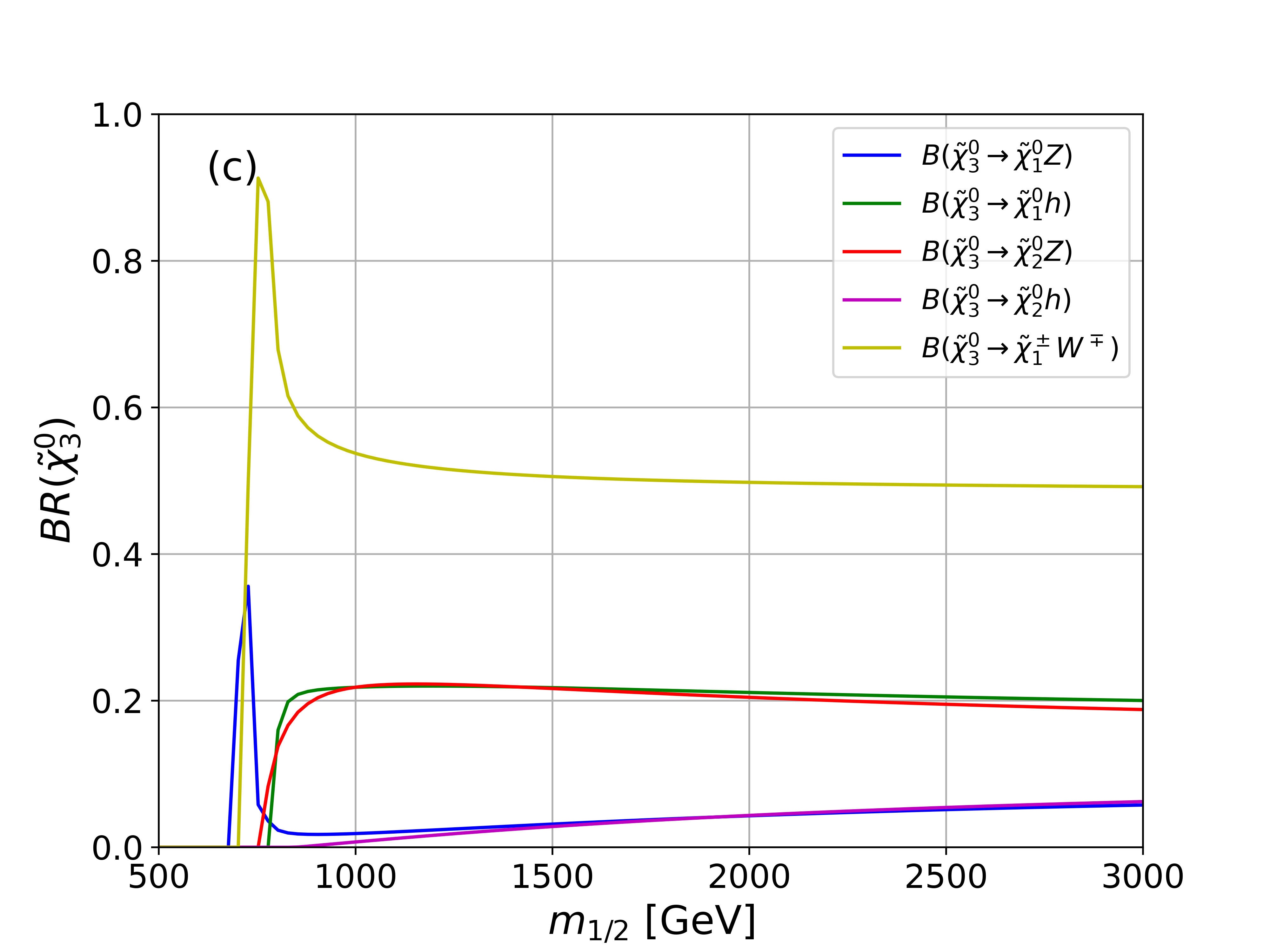}}
    \caption{Branching fractions from natural SUSY versus $m_{1/2}$ for the
      model line defined by Eq.~(\ref{eqn:mline}) for ({\it a})~the charged
      wino-like state, $\tw_2$, ({\it b})~the neutral wino-like state,
      $\tz_4$, and ({\it c}) the bino-like state $\tz_3$. We show the
      branching fractions only for two-body decay channels, all of which are
      closed for small values of $m_{1/2}$ in frame ({\it c}).  
      \label{fig:bfs}}
\end{figure}

LHC signatures for wino production depend on how these decay. The
branching fractions, within natural SUSY, for decays of the charged
and the neutral wino-like states are shown versus $m_{1/2}$ in
Fig.~\ref{fig:bfs} frames ({\it a}) and ({\it b}), respectively.
We obtain the brnching fractions from ISAJET.
As in Fig.~\ref{fig:prod}, the other parameters are fixed by
Eq.~(\ref{eqn:mline}). We see that for $m_{1/2}\agt 1$~TeV where phase
space effects are unimportant for the analysis of wino decays,
$B(\tw_2\to W\tz_{1,2}): B(\tw_2 \to h\tw_1):B(\tw_2\to Z\tw_1)\simeq
2:1:1$,
with a very small fraction of the $\tw_2$
decaying via the dynamically and kinematically suppressed decay to the
bino. Here, we sum over the decays to the neutral higginos since as
mentioned previously, we do not attempt to identify the soft visible
decay products of the higgsinos. Likewise, from frame ({\it b}) we see
that even for neutral winos, $B(\tz_4\to W^\mp\tw_1):B(\tz_4\to
h\tz_{1,2}):B(\tz_4\to Z\tz_{1,2}) =2:1:1$ as long as these are heavy,
while the decay to the bino is again strongly suppressed. The reason
for this simple 2:1:1 pattern of charged and neutral branching ratios
has been explained in Ref.~\cite{Baer:2017gzf} and we will not repeat
it here.

Although not germane to the considerations of this paper, for
completeness we show the branching fractions for the {\em two-body
  decays} of the bino-like $\tz_3$ state in Fig.~\ref{fig:bfs}{\it   c}.
For small values of $m_{1/2}$ in the left-most part of the
frame, the $\tz_3$ is too light to decay to higgsinos together with an
on-shell vector boson and so decays via three-body modes. The first
two-body modes to become accessible are
$\tz_3\to\tz_1Z$ and
$\tz_3\to\tw_1 W$. Indeed in the region where $m_{1/2} \sim
700-800$~GeV where the two-body decays turn on, $M_1$ is comparable to
$\mu$ and the lighter states are well-mixed binos and higgsinos, with
only the heaviest states being wino-like. For values of $m_{1/2}$
larger than $\sim 0.9-1$~TeV, our expectation that the mass
eigenstates are dominantly higgsino-, bino- or wino-like, and we see
once again the simple 2:1:1 decay pattern for decays to $W$, $Z$ and
$h$ bosons, for more or less the same reason as for the decay of the
neutral wino states \cite{Baer:2017gzf}.

\section{Wino discovery channels}
\label{sec:channels}

We now turn our attention to the potential of HL-LHC ($\sqrt{s}=14$ TeV
with 3000 fb$^{-1}$) for probing wino pair production in the context
of natural SUSY. We use Isajet to first construct a SUSY Les Houches
Accord (SLHA) file\cite{Skands:2003cj} for any natural SUSY
parameter-space point and feed this into Pythia\cite{Sjostrand:2006za} to
generate signal events. We also use Pythia to generate the various $2\to
2$ background (BG) processes.  For $2\to 3$ background processes, we use
Madgraph\cite{Alwall:2011uj}, coupled with Pythia.  For our computation
of SM backgrounds to the gaugino signal, we include parton level
production of $t\bar{t}$, $t\bar{t}V$, $ht\bar{t}$, $hh$, $hhV$,
$V+jets$, $Vh$, $VV$, and $VVV$ final states to evaluate SM
backgrounds. Specifically, we normalize the SUSY cross sections to their
NLO values obtained from Prospino. For the most important SM backgrounds
we normalize the cross sections to their values at the NLO
level, or better when available.
The NNLO/NNLL $t\bar{t}$ cross section is normalized to
985.7~pb,\footnote{This is taken from
  {\it https://twiki.cern.ch/twiki/bin/view/LHCPhysics/TtbarNNLO} where
  references to the literature for the calculation may also be found.}
the cross sections for $t\bar{t}V$ production are from
Ref.\cite{LHCHiggsCrossSectionWorkingGroup:2016ypw}, $V+j$ and $V
b\bar{b}$ cross sections are 
calculated using the $K$-factor from the ratio of NLO and LO cross
sections from MadGraph with parton jets defined using the anti-$k_T$
algorithm with $p_{Tj}>25$~GeV and $\Delta R=0.4$, and finally $VV$
cross sections are normalized using the results in
Ref.\cite{Campbell:2011bn}. The remaining backgrounds which are
frequently orders of magnitude smaller are included at leading
order. We use the Delphes code for detector simulation in our
analysis\cite{deFavereau:2013fsa}.

Since our signal consists of combinations of high transverse momentum
$W, Z$ and $h$ bosons decaying leptonically or hadronically, we focus on
hard leptons and jets in the central part of the detector. With this in
mind, we require isolated electrons and muons to satisfy,
\begin{itemize}
\item $p_T(e)> 20$ ~GeV, $|\eta_e| < 2.47$, with $P_{TRatio}< 0.1$,  and 

\item $p_T(\mu)> 25$ ~GeV,  $|\eta_\mu| < 2.5$ with $p_{TRatio}<0.2$,
\end{itemize}
where $P_{TRatio}$ is defined as the ratio of the scalar sum of the
transverse momentum of objects (defined in the default Delphes
configuration card for HL-LHC simulation) in a $\Delta R=0.3$ cone about
the lepton.

We construct small radius (SR) jets using an anti-$k_T$ jet algorithm
and require,
\begin{itemize}
\item  $p_T(SRj)>25$~GeV with a cone size $R \le 0.4$ and
$|\eta(SRj)|<4.5$. 
\end{itemize}
The jet is labeled as a $b$-jet if, in addition, it
satisfies,
\begin{itemize}
\item $|\eta_j| < 2.4$ and is tagged as a $b$-jet by Delphes. 
\end{itemize}
Since our aim is to also identify hadronically decaying high $p_T$ gauge
and Higgs bosons,\footnote{The analyses of electroweak gaugino signals
  at hadron colliders before the LHC focused on the leptonic decays of
  the gauginos because backgrounds to the hadronic signals were thought
  to be large. Subsequent developments in our understanding of jet
  substructure now allow us to zero in on the signal from hadronic
  decays of boosted, heavy daughters of heavy gauginos without being
  completely overwhelmed by QCD radiation. Gaugino searches at even the
  Tevatron were confined to gaugino mass ranges where the gaugino
  decayed via three-body decays, or where the daughter bosons from
  gaugino decay were not particularly boosted. An additional factor that
  may have precluded earlier exploration of hadronic decays of gauginos
  is that the magnitiude of $\eslt$ from hadronic mismeasurements scales
  as square root of the total scalar $E_T$ in the event, and so is
  fractionally less relevant compared to the physics $\eslt$ at higher
  masses probed at the LHC.}  we construct large radius (LR) jets using
an anti-$k_T$ jet algorithm with a cone $R < 1.5$ , and require
\begin{itemize}
\item  $p_T(LRj)>300$~GeV and $\eta(LRj) < 2$. 
\end{itemize}

We use the identified leptons, SR jets and LR jets to construct $W$,$Z$
and $h$ candidates as  follows.\\

 {\it Leptonically decaying $Z$}:

A pair of opposite sign (OS) leptons with the same flavour (SF)
satisfying 80~GeV $< m(\ell\ell)<$~100~GeV is identified as a
candidate $Z$ boson. \\

{\it Hadronically decaying $W$}:

 A LR jet with trimmed mass\cite{Krohn_jet_trim} $60\ {\rm GeV} <
  m_J < 90$ GeV is identified as a candidate $W$ boson.\\

{\it Hadronically decaying $Z$}:

Either (or both)
\begin{enumerate}
\item[a)] two small radius signal $b$-jets that have an invariant mass
  $80\ {\rm GeV} < m(bb) < 100$ GeV,

or
\item[b)]a LR jet with trimmed mass $70\ {\rm GeV} < m_J < 100$ GeV,
\end{enumerate}
defines  a candidate $Z$ boson. 
We do not attempt to distinguish between LR jets from $W$
and $Z$ bosons, but simply classify LR jets with a trimmed mass between 
 $60\ {\rm GeV} < m_J < 100$ GeV as a vector boson.  \\

 {\it Higgs bosons}: 

Either (or both)
\begin{enumerate}
\item[a)] two small radius signal $b$-jets that have an invariant mass $100\ {\rm GeV} < m(bb) < 135$ GeV,

or
\item[b)] LR jet with trimmed mass $100\ {\rm GeV} < m_J < 135$ GeV, and
  at least one small radius ($R<0.4$) signal $b$-jet within the cone
  radius of the LR jet,
\end{enumerate}
define a candidate  SM-like Higgs boson, $h$. \\

Having discussed how we identify candidate hadronically decaying
$W$, $Z$ and $h$ daughters of winos (these have the largest branching
fractions and so lead to the largest signal rates) we proceed to
classify events into eight channels: 
\begin{enumerate}
\item $Z(\to \ell^+\ell^-) B+\eslt$, 

\item $h/Z(\to bb)B+\eslt$, 

\item $BB+\eslt$, 

\item $\ell h+\eslt$, 

\item $\ell B_{W/Z}+\eslt$, 

\item $Z(\to \ell^+\ell^-) +\eslt$,  

\item $h/Z(\to bb)+\eslt$, and

\item $\ell^\pm\ell^\pm +\eslt$ events from $q\bar{q'} \to
  \widetilde{W}^\pm(\to W^{\pm}\tilde{h}^0)\widetilde{W}^0(\to
  W^{\pm}\tilde{h}^{\mp})$, where the $W$ bosons decay leptonically and
  the decay products of higgsinos are soft so that these events have
  hadronic activity only from QCD radiation
  \cite{Baer:2013yha,Baer:2017gzf}.
\end{enumerate}
Here, $B$ (for boson) means any hadronically decayed $W$, $Z$ or $h$
boson as defined above, while $B_{W/Z}$ refers to hadronically
decaying $W$ and $Z$ bosons identified as LR jets with
60~GeV~$<m_J<$~100~GeV. It may appear that the same event may be
classified in more than one channel; {\it e.g.} an event with a Higgs
boson $h\to bb$ produced in association with a hadronically decaying
$W$ may be classified in both the $h/Z(\to bb)B+\eslt$ as well as the
$BB+\eslt$ channels. We have ensured this is not the case.
If an event satisfies the criteria in more than one channel, we 
classify it to be in the channel that appears first on the
list, and do not count it again in any of the subsequent channels. The
fact that these channels are non-overlapping will be important when we
combine them to project the significance of the signal. These channels
are defined as follows.
\begin{enumerate}
    \item $Z(\to \ell^+\ell^-) B+\eslt$:
        \begin{itemize}
            \item Exactly one pair of OS/SF  leptons;
            \item $80\ {\rm GeV} < m(\ell\ell) < 100$~GeV;
            \item One hadronically decayed $W/Z/h$;
            \item If the hadronically decayed $W/Z/h$ is tagged with a
              LR jet, the two  leptons should be outside the cone
              radius of this LR jet $\Delta R(J, \ell) > 1.5$.
        \end{itemize}

    \item $h/Z(\to bb)B+\eslt$:
         Events that fail the previous classification criterion, but satisfy,
        \begin{itemize}
            \item No isolated leptons;
            \item At least two  $b$ jets;
            \item At least one pair of $b$ jets has an invariant mass of
              $80\ {\rm GeV} < m(bb) < 135$ GeV;
            \item Besides the $b$ pair that reconstructs $h/Z$, one other
              hadronically decayed $W/Z/h$;
            \item If the hadronically decaying boson is tagged with a
              LR jet, the two $b$-jets that reconstruct $h/Z$ should
              be outside the cone radius of the LR jet $\Delta R(J,
              b) > 1.5$, with no other $b$ jets are allowed outside
              the cone radius of the LR jet.
        \end{itemize}

    \item $BB+\eslt$:
        Events that fail all the above classification criteria, but contain,
        \begin{itemize}
            \item two hadronically decayed $W/Z/h$, each being tagged
              with a LR jet.
        \end{itemize}

    \item $\ell h+\eslt$:
        Events that fail all the above classification criteria, but contain,
        \begin{itemize}
            \item at least one isolated lepton;\footnote{We found
              requiring at least one isolated lepton proved better than
              exactly one isolated lepton because $\sim 10$\% of signal
              events contained additional leptons, presumably from the
              decays of the daughter higgsinos.}
            \item exactly one Higgs boson $h$;
            \item if the Higgs is tagged with a LR jet, the lepton
              candidates should be outside the cone radius of this LR
              jet $\Delta R(J, \ell) > 1.5$;
            \item if there are two or more  lepton candidates, the
              one that minimizes the magnitude  of
              $\vec{\eslt}+\vec{p}_T(h)+\vec{p}_T(\ell)$ is chosen as
              the  lepton.
        \end{itemize}

    \item $\ell B_{W/Z}+\eslt$: 
        Events that fail all the above classification criteria, but contain,
        \begin{itemize}
            \item at least one isolated lepton;
            \item exactly one hadronically decaying W/Z boson tagged
              as a LR jet;
            \item the lepton candidates should be outside the cone
              radius of the signal LR jet $\Delta R(J, \ell) > 1.5$;
            \item if there are two or more lepton candidates, the
              one that minimizes the magnitude of
              $\vec{\eslt}+\vec{p}_T(W/Z)+\vec{p}_T(\ell)$ is chosen as
             the  lepton.
        \end{itemize}

    \item $Z(\to \ell^+\ell^-) +\eslt$:
        Event that fail all the above classification criteria, but contain,
        \begin{itemize}
            \item exactly one pair of OS/SF  leptons;
            \item $80\ {\rm GeV} < m(\ell\ell) < 100$ GeV.
        \end{itemize}

    \item $h/Z(\to bb)+\eslt$:
        Events that fail all the above classification criteria, but contain,
        \begin{itemize}
            \item at least two  $b$ jets;
            \item exactly one pair of $b$ jets has an invariant mass of
              $80\ {\rm GeV} < m(bb) < 135$ GeV.
     \end{itemize}

\item $\ell^\pm\ell^\pm+\eslt$:
Events that fail all the above classification criteria but contain,      
\begin{itemize}

\item Exactly two same-sign leptons with large $\eslt$;
\item No tagged $b$-jets.

\end{itemize}

    \end{enumerate}

Having decided on the various channels for the wino search, we now
proceed with the analysis of the signal in each of these
channels. Toward this end, we have examined several signal and SM
background distributions to develop analysis cuts that serve to
enhance the gaugino signal in each of the eight channels. We do not
show these distributions here, instead only display the additional
channel-by-channel cuts that we use to assess the observability of the
gaugino signal over SM backgrounds. For each channel with two tagged
bosons (including the $\ell h+\eslt$ and $\ell B_{W/Z}+\eslt$
channels) we show $m_{T2}$\cite{Barr:2003rg}
distributions after analysis cuts, while
for the channels with just a single tagged boson we show corresponding
$m_T$ distributions that we use for subsequent statistical analysis of
the observability of the signal. For the same sign dilepton plus
$\eslt$ channel, we show instead the distribution of $L_T\equiv
|p_T(\ell_1)|+|p_T(\ell_2)|+|\eslt|$. Our goal is to examine
whether these distributions are significantly modified from SM
expectation due to the presence of a signal.

\subsection{$Z(\to \ell^+\ell^-) B+\eslt$ channel}

We begin our investigation with the $Z(\to \ell^+\ell^-) B+\eslt$
channel. Upon analyzing various kinematics distributions, we require
additional analysis cuts:

\begin{itemize}
    \item $\eslt > 350$ GeV;
    \item $max[m_T(Z(\ell^+\ell^-), \eslt), m_T(B, \eslt)] > 1000$ GeV;
    \item $min[m_T(b, \eslt)] > 175$ GeV, where $b$ loops over all $b$-jets;
    \item $\Delta R(\ell^+, \ell^-) < 0.8$;
    \item $min[\Delta\phi (J, \eslt)] > 35^\circ$, where $J$ loops over
      all LR jets in the event, no matter whether these have been tagged
      as a boson or not.
\end{itemize}

\begin{figure}[htb!]
\begin{center}
  \includegraphics[height=0.4\textheight]{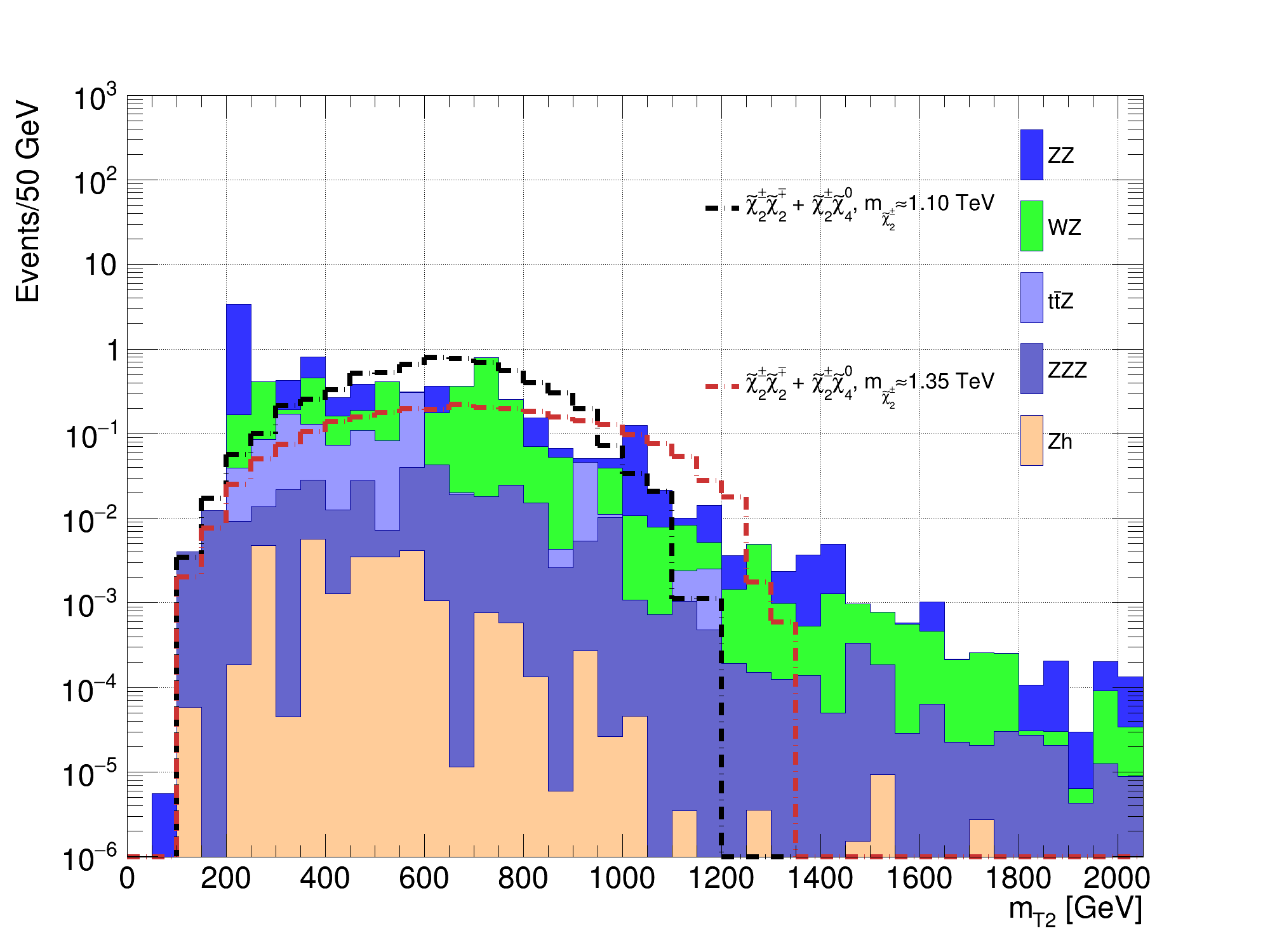}
  \caption{The distribution of $m_{T2}(Z(\ell^+\ell^-) B,\eslt)$ after the
    analysis cuts detailed in the text for the wino signal for
    $m_{1/2}=1.3$ and 1.6~TeV, with other parameters fixed as
    in Eq.~(\ref{eqn:mline}) is shown by the hollow histograms. The
    corresponding background distributions are shown by the filled
    histograms. The background histograms are not stacked.  
  \label{fig:mT2_2l1J}}
\end{center}
\end{figure}

The $m_{T2}$ distribution, after these cuts, is shown in
Fig. \ref{fig:mT2_2l1J} for both the signal as well as for the leading
SM backgrounds.  We show the signal histograms for $m_{1/2}=1.3 and
1.6$~TeV (corresponding to a charged wino mass of 1.1 and 1.35~TeV,
respectively) along the model line defined by Eq.~(\ref{eqn:mline}). The
primary backgrounds are depicted by the filled histograms and mainly
come from $WZ$ and $ZZ$ production. The lower $m_{T2}$ region is
dominated by the background.  The signal begins to emerge from the
background at increased $m_{T2}$ values determined by the parent gaugino
masses until it cuts off at high values of $m_{T2}$. Our goal is to
examine whether the wino signal sufficiently distorts the expectation of
the event rate from SM expectation so that one can claim a discovery of
new physics for the chosen value of $m_{1/2}$ after combining (see
Sec.~\ref{sec:reach} below) the eight different channels that we
analyse.  If instead the signal is too small to cause a sufficient
deviation from SM expectations, the corresponding value of $m_{1/2}$ can
be excluded.

\subsection{$h/Z(\to bb)B+\eslt$ channel}
For wino searches via this channel, after analyzing  various
distributions, we require additional analysis cuts:

\begin{itemize}
    \item $\eslt > 450$ GeV;
    \item No jet in the event is tagged as $\tau$ by Delphes;
    \item $max[m_T(h/Z(bb), \eslt), m_T(B, \eslt)] > 1100$ GeV;
    \item $min[m_T(b, \eslt)] > 175$ GeV, where $b$ loops over all $b$-jets,
    \item $min[\Delta\phi (J, \eslt)] > 35^\circ$, where $J$ loops over
      all LR jets in the event, whether or not these have been tagged
      as a $W$, $Z$ or $h$ boson;
    \item No LR jets in the event should have a trimmed mass in the mass
      range of top, so $m_J \notin (135, 185)$ GeV.
\end{itemize}

\begin{figure}[htb!]
\begin{center}
  \includegraphics[height=0.4\textheight]{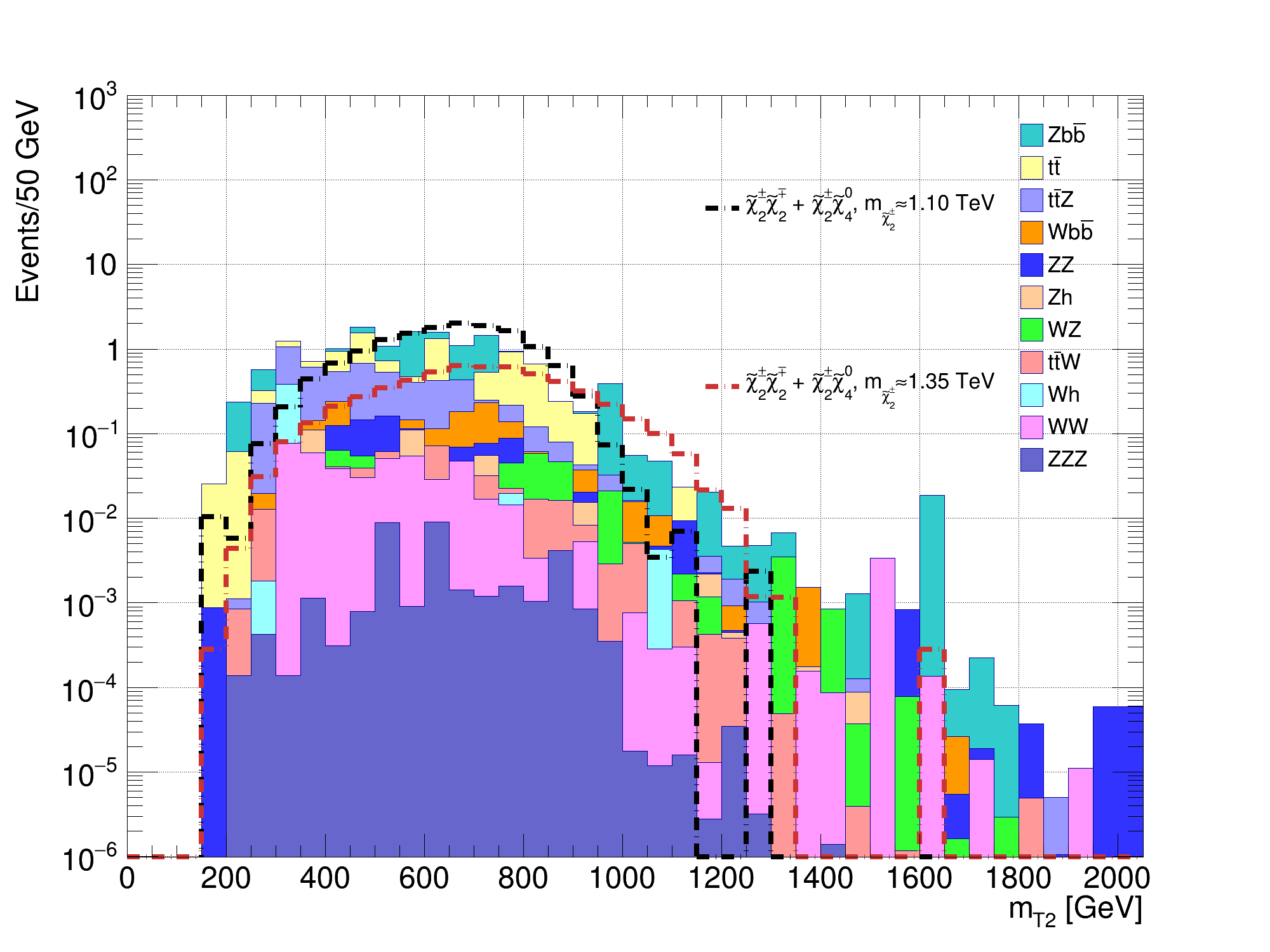}
  \caption{The distribution of  $m_{T2}(h/Z(bb) B,\eslt)$ after the
    analysis cuts detailed in the text for the wino signal for
    $m_{1/2}=1.3$ and 1.6~TeV, with other parameters fixed as
    in Eq.~(\ref{eqn:mline}) is shown by the hollow histograms. The
    corresponding background distributions are shown by the filled
    histograms. The background histograms are not stacked.   
  \label{fig:mT2_2b1J}}
\end{center}
\end{figure}

The resulting $m_{T2}$ distributions are displayed in
Fig. \ref{fig:mT2_2b1J} for both the two signal cases (hollow
histograms) shown in the last figure as well as leading SM backgrounds
(solid histograms). In this channel, the largest backgrounds are from
$t\bar{t}$ and $Z+b\bar{b}$ production. Once again, we see that the
signal distributions have broad peaks and cut off at $m_{T2}$ values
determined by the wino mass, while the background is largely a broad
continuum. The signal histograms distinctly rise above the background at
large $m_{T2}$ values before their kinematic cut-off. 

\subsection{$BB +\eslt$ channel}
Next, we examined various distributions for signal and background in the
$BB +\eslt$ channel to arrive at the following analysis cuts:
\begin{itemize}
    \item $\eslt > 200$ GeV;
    \item No jet in the event is tagged as $\tau$ by Delphes;
    \item $max[m_T(B_1, \eslt), m_T(B_2, \eslt)] > 1000$ GeV;
    \item $min[m_T(b, \eslt)] > 175$ GeV, where $b$ loops over all
      $b$-jets;
    \item $min[\Delta\phi (J, \eslt)] > 35^\circ$, where $J$ loops over
      all LR jets in the event whether they have been tagged
      as a boson;
    \item No LR jet in the event should have a trimmed mass in the mass
      range of top, so $m_J \notin (135, 185)$ GeV.
\end{itemize}

\begin{figure}[htb!]
\begin{center}
  \includegraphics[height=0.4\textheight]{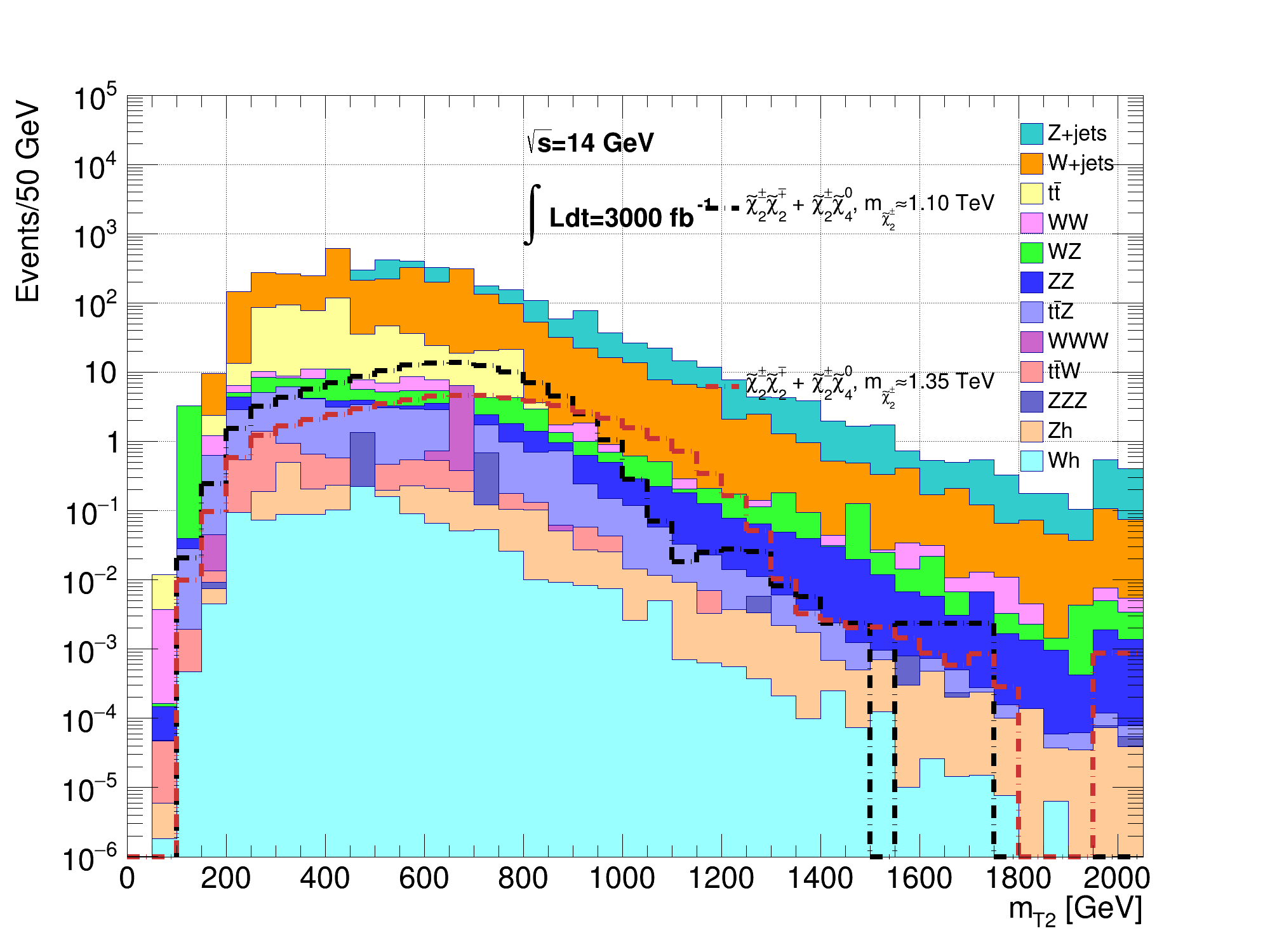}
  \caption{The distribution of $m_{T2}(BB,\eslt)$ after the
    analysis cuts detailed in the text for the wino signal for
    $m_{1/2}=1.3$ and 1.6~TeV, with other parameters fixed as
    in Eq.~(\ref{eqn:mline}) is shown by the hollow histograms. The
    corresponding background distributions are shown by the filled
    histograms. The background histograms are not stacked.   
  \label{fig:mT2_2J}}
\end{center}
\end{figure}

The resulting signal and background $m_{T2}$ distributions are shown
in Fig. \ref{fig:mT2_2J}, again for the same signal cases as
before. As in previous figures, the signal distribution is bounded
above by the wino mass. In this channel,\footnote{Bear in mind that
$Z(\to\ell\bar{\ell})B+\eslt$ events and $h/Z(\to bb) B+\eslt$ events
which have been included in previous channels are not counted in this
channel.} however, the enormous $W/Z + jets$ and also the $t\bar{t}$
backgrounds completely overwhelm the signal even after selection
cuts. We might think that this channel will make a negligible
contribution to the significance of the wino signal at the HL-LHC.
Notice, however, that the signal cross sections as well as the
backgrounds in this channel are an order of magnitude larger than for
other channels discussed, so that the naive measure ``$S/\sqrt{B}$''
(over a few optimally chosen bins) might make this channel
competitive. Systematic uncertainties may change this picture though.

\subsection{$\ell^\pm h +\eslt$ channel}

In this channel our intent is to target events where one of the winos
decays into a leptonically decaying $W$ boson,  while
the other wino decays to the light Higgs boson. After examining various
distributions we further require:
\begin{itemize}
    \item $\eslt > 450$ GeV;
    \item No jet in the event is tagged as $\tau$ by Delphes;
    \item $max[m_T(\ell, \eslt), m_T(h, \eslt)] > 1100$ GeV;
    \item $min[m_T(b, \eslt)] > 175$ GeV, where $b$ loops over all $b$-jets;
    \item $\Delta\phi (h, \eslt) > 115^\circ$;
    \item $min[\Delta\phi (J, \eslt)] > 65^\circ$, where $J$ loops over
      all LR jets in the event, whether or not these have been tagged
      as a boson;
    \item No LR jets in the event should have a trimmed mass in the mass
      range of top, so $m_J \notin (135, 185)$ GeV.
\end{itemize}

\begin{figure}[htb!]
\begin{center}
  \includegraphics[height=0.4\textheight]{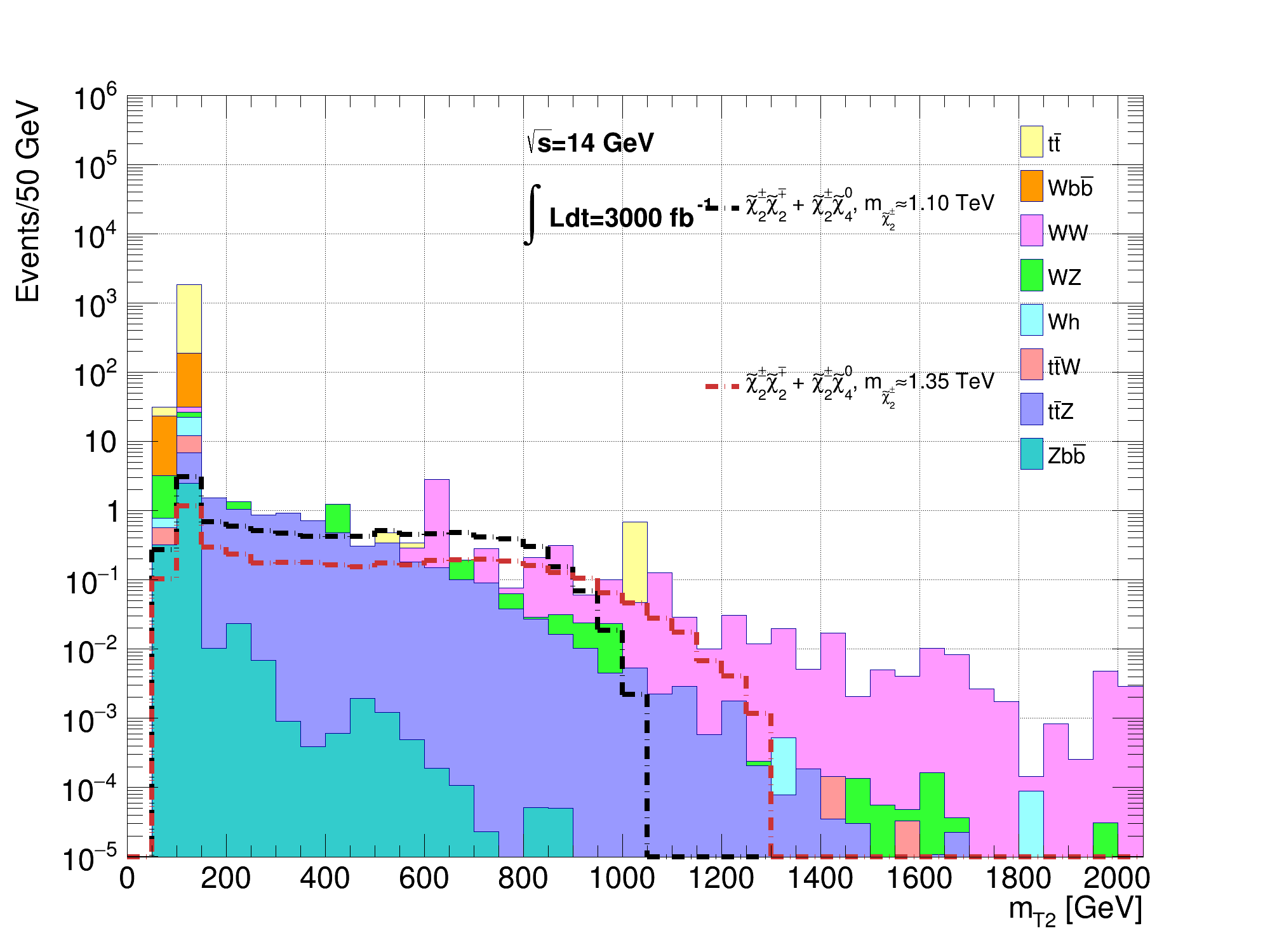}
  \caption{The distribution of   $m_{T2}(\ell h,\eslt)$ after the
    analysis cuts detailed in the text for the wino signal for
    $m_{1/2}=1.3$ and 1.6~TeV, with other parameters fixed as
    in Eq.~(\ref{eqn:mline}) is shown by the hollow histograms. The
    corresponding background distributions are shown by the filled
    histograms. The background histograms are not stacked.
  \label{fig:mT2_1l1h}}
\end{center}
\end{figure}

The $m_{T2}$ distributions after these  cuts are then shown in
Fig. \ref{fig:mT2_1l1h} for the two signal cases as well as for
various SM backgrounds. SM processes involving $W$ boson production, either
directly from $VV$ pair production or from decays of top quarks,
constitute the dominant backgrounds.
The signal distributions are again clearly bounded by the wino
mass. Somewhat surprising is the long background tail from SM $WW$
production because the LR jet from the hadronic decay of the $W$-boson
would not be expected to have the trimmed mass in the 100-135~GeV range,
or for that matter include a SR $b$-jet.
However, $b$ quarks from QCD radiation (or
jets mistagged as a $b$) could combine with either a tau jet or hadronic
decay products of the $W$ that is {\em not}
the parent of the lepton to push the LR jet mass into the higher range,
causing it to be  tagged as an $h$. The signal cross sections are
comparable in magnitude to the background cross sections and we may
anticipate that this channel will contribute to the significance of
the signal at the HL-LHC.

\subsection{$\ell B_{W/Z}+\eslt$ channel}

This channel is designed to examine events where one wino decays into a
leptonically decaying $W$ boson and the other decays into a hadronically
decaying $W/Z$ boson tagged as a LR jet, but not a $Z$ boson tagged via
two $b$-jets reconstructing the $Z$. Upon examination of various
distributions, we further require:
\begin{itemize}
    \item $\eslt > 500$ GeV;
    \item No jet in the event is tagged as $\tau$ by Delphes;
    \item $max[m_T(\ell, \eslt), m_T(B_{W/Z}, \eslt)] > 1000$ GeV;
    \item $min[m_T(b, \eslt)] > 175$ GeV, where $b$ loops over all $b$-jets;
    \item $\Delta\phi (B_{W/Z}, \eslt) > 105^\circ$;
    \item $min[\Delta\phi (J, \eslt)] > 15^\circ$, where $J$ loops over
      all LR jets in the event, whether or not these have been tagged
      as a boson.
\end{itemize}

\begin{figure}[htb!]
\begin{center}
  \includegraphics[height=0.4\textheight]{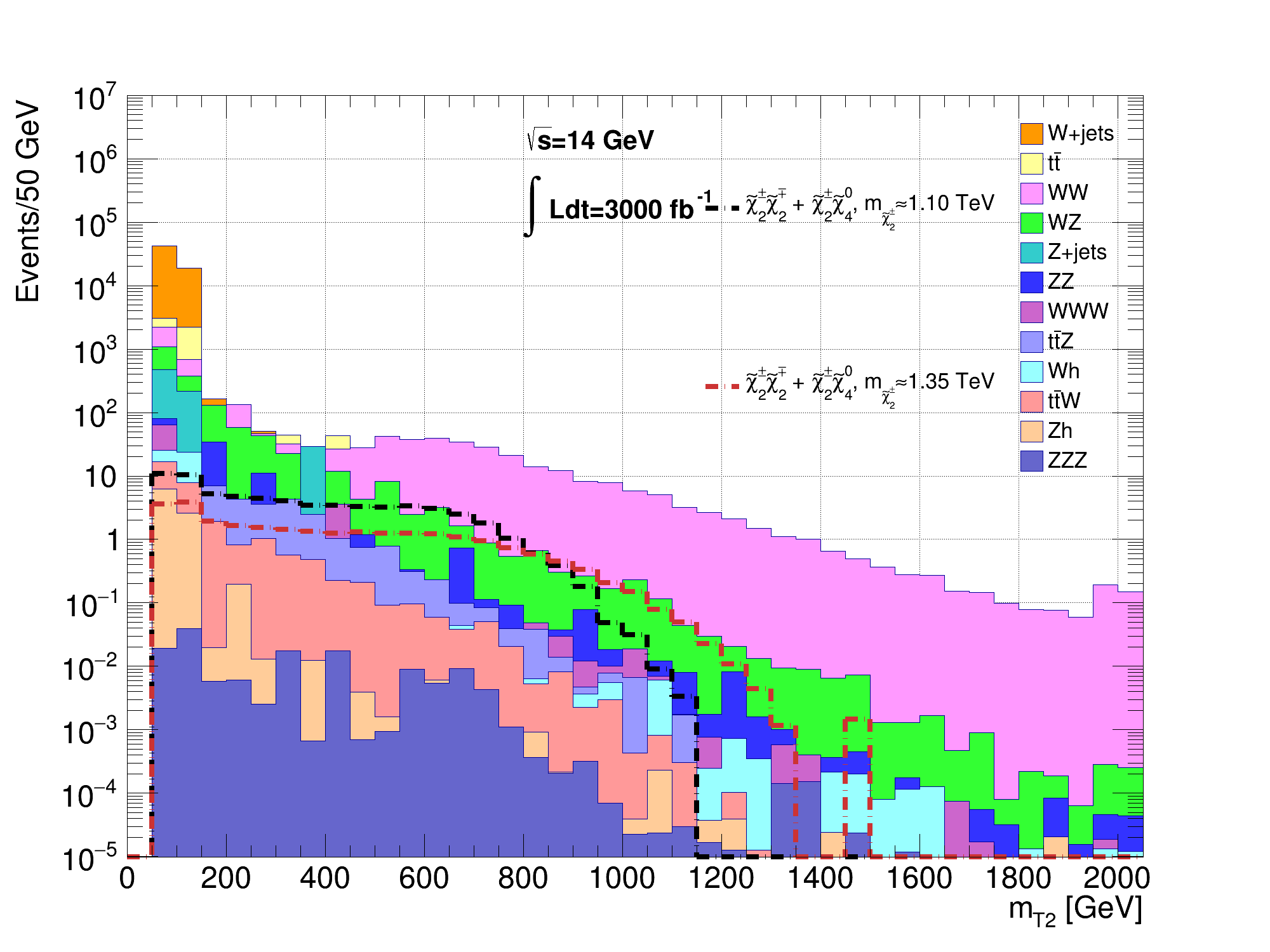}
  \caption{The distribution of $m_{T2}(\ell B_{W/Z},\eslt)$ after the
    analysis cuts detailed in the text for the wino signal for
    $m_{1/2}=1.3$ and 1.6~TeV, with other parameters fixed as
    in Eq.~(\ref{eqn:mline}) is shown by the hollow histograms. The
    corresponding background distributions are shown by the filled
    histograms. The background histograms are not stacked.
  \label{fig:mT2_1l1J}}
\end{center}
\end{figure}

The $m_{T2} $ distributions for the two signal cases as well as for
various SM backgrounds are shown in Fig.~\ref{fig:mT2_1l1J}. As may
have been anticipated, SM processes involving $WW$ and $WZ$ pair
production are the dominant background source except for the smallest
values of $m_{T2}$ where $Wj$ production dominates. While the
distribution of events from $WW$ and $WZ$ production do exhibit a peak
at $m_{T2}\alt 100-150$~GeV, the long tail extending to TeV values of
$m_{T2}$ where we expect the signal to reside may seem somewhat
surprising. We have checked that events with $m_T(\ell,\eslt) < 100$~GeV
are essentially all in the low $m_{T2}$ peak so that imposing a cut on
this does not allow the signal to stand out above the long tail in the
$WW$ background.\footnote{Events where one $W$ decays leptonically and
the other boson decays hadronically would be expected to satisfy
$m_T(\ell,\eslt) < 100$~GeV if the single neutrino from the leptonic
decay of $W$ is the primary source of $\eslt$.} We have further
checked that in most of the events in the tail contain two neutrinos,
the second neutrino (for the most part) coming from $W\to\tau\nu$ or
from $W\to s c(\to \nu)$ decays. The long tail presumably comes
from the fact that hard QCD radiation forms part of the LR jet, {\it
  i.e.}  the $B$ is not entirely composed of the decay products of the
second $W$. As in the $BB+\eslt$ channel studied above, it appears
that the backgrounds are one and a half orders of magnitude higher than
the signal, but bear in mind that the signal event rate is also higher
than in many of the channels where signal and background were comparable in
a range of mass bins.

\subsection{$Z(\to \ell^+\ell^-) +\eslt$ channel}

This channel is designed to catch events where both winos decay to a $Z$
boson, one of which decays leptonically, and the other invisibly to
neutrinos. One expects enormous $\eslt$ in these events. 
After exploring several distributions, we further require:
\begin{itemize}
    \item $\eslt > 750$ GeV,
    \item $L_T > 1550$ GeV, where $L_T$ is defined to be the scalar sum of the $p_T$ of all jets and leptons, and $\eslt$ in the event,
    \item $m_{CT} > 100$ GeV, where $m_{CT} = \sqrt{2
      p_T(\ell^+)p_T(\ell^-)(1+\cos{(\Delta\phi(\ell^+, \ell^-)))}}$
\end{itemize}

\begin{figure}[htb!]
\begin{center}
  \includegraphics[height=0.4\textheight]{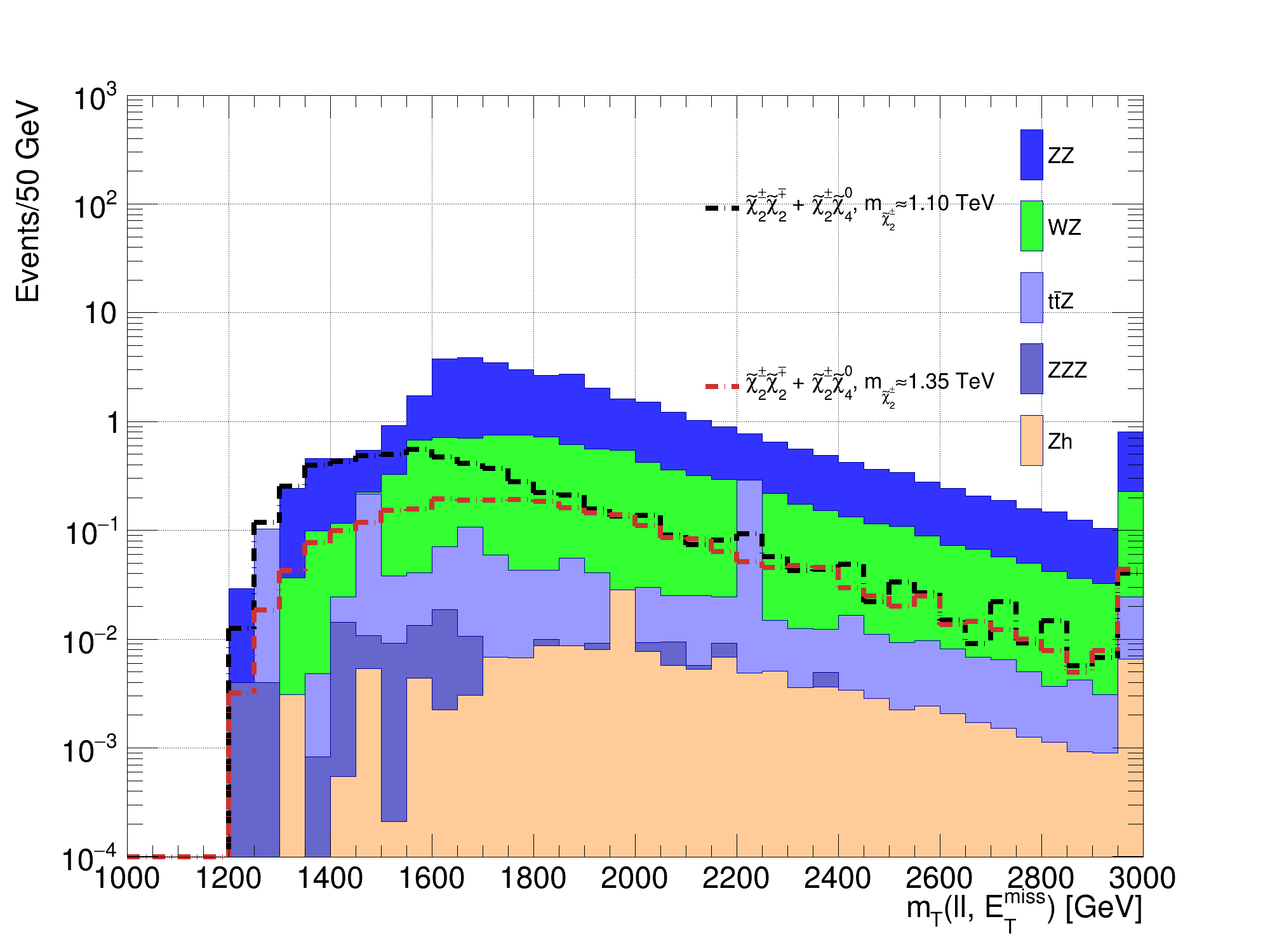}
  \caption{The distribution of  $m_T(\ell^+\ell^-,\eslt)$ after the
    analysis cuts detailed in the text for the wino signal for
    $m_{1/2}=1.3$ and 1.6~TeV, with other parameters fixed as
    in Eq.~(\ref{eqn:mline}) is shown by the hollow histograms. The
    corresponding background distributions are shown by the filled
    histograms. The background histograms are not stacked.
  \label{fig:mT_2lOS}}
\end{center}
\end{figure}

Since only one boson is reconstructed in this channel, we show the
distribution of cluster transverse mass\cite{Barger:1983jx} $m_T(\ell^+\ell^-,\eslt)$ for this
channel in Fig. \ref{fig:mT_2lOS} for the two signal cases and for
various SM backgrounds. Not surprisingly,
$Z(\to\ell^+\ell^-)Z(\to\nu\bar{\nu})$ dominates the SM background
followed by $Z(\to\ell^+\ell^-) W$ production where the $W$ decays to an
$e,\mu$ or $\tau$ that is missed in the detector. Although the
backgrounds are large for $m_T\agt 1.5$~TeV, the backgrounds are
comparable to the wino signal over a significant range, and it seems
possible that a distortion of this distribution due to the presence of a
signal may contribute to its overall significance when the various
channels are combined.

\subsection{$h/Z(\to 2b) +\eslt$ channel}

This channel is designed to catch events where one wino decays to an
$h$ or $Z$ boson tagged by two SR $b$-jets with an invariant mass consistent
with $m_Z$ or $m_h$, and the other wino decays to a $Z$ that is
essentially invisible. There would, of course, be contributions to this
channel where the boson on the other side fails to be tagged, {\it e.g}
it is a $W$ decaying via $W\to\tau\nu$, and the hadronically decaying
tau is not identified.

For this channel, we require
\begin{itemize}
    \item $\eslt > 850$ GeV,
    \item No jet in the event is tagged as $\tau$ by Delphes,
    \item $min[m_T(b, \eslt)] > 175$ GeV, where $b$ loops over all $b$-jets,
    \item $min[\Delta\phi (b, \eslt)] > 85^\circ$, where $b$ loops over all $b$-jets,
    \item No LR jets in the event should have a trimmed mass in the
      mass range of top, so $m_J \notin (135, 185)$ GeV.
\end{itemize}

\begin{figure}[htb!]
\begin{center}
  \includegraphics[height=0.4\textheight]{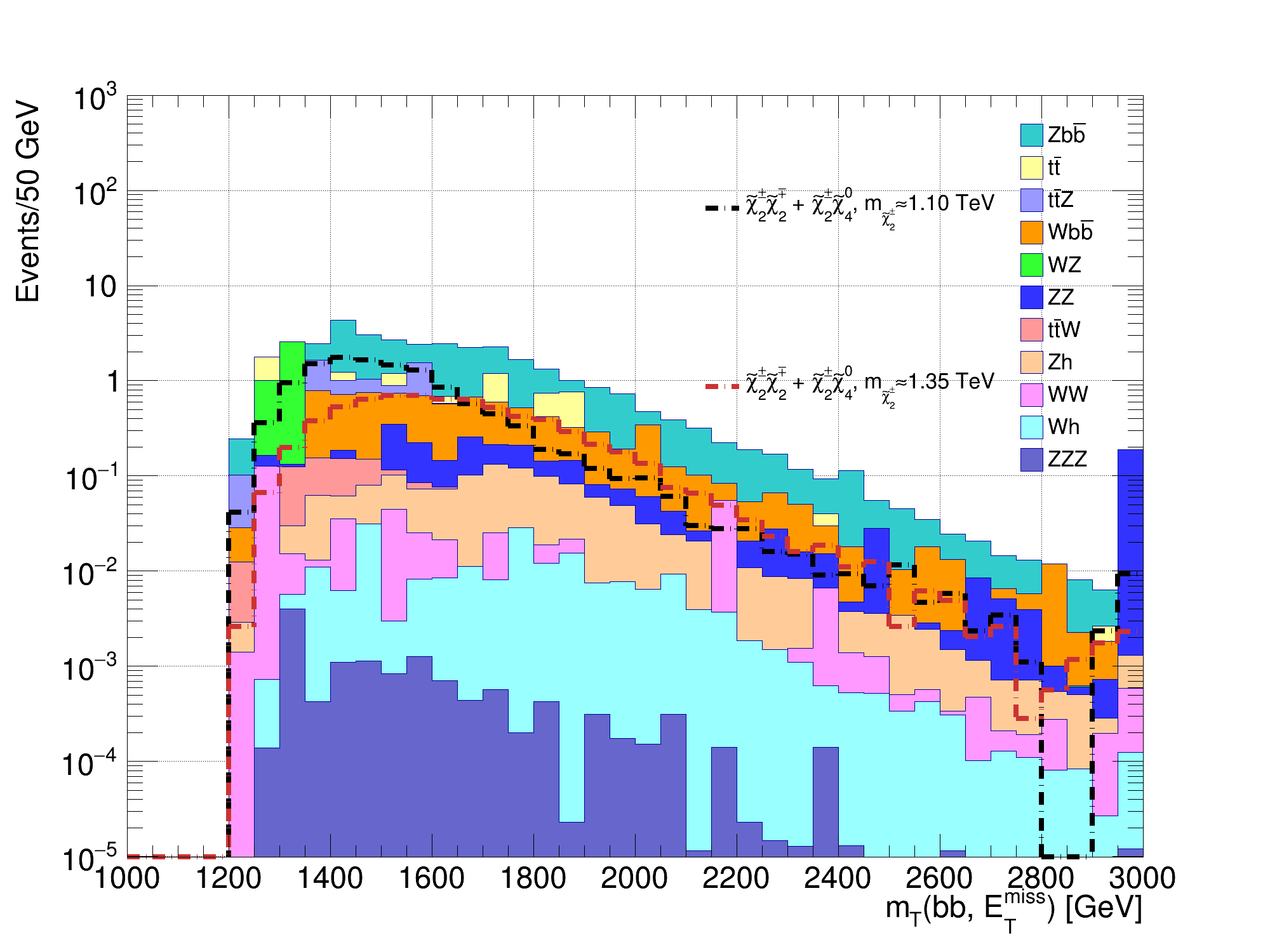}
  \caption{The distribution of  $m_T(bb,\eslt)$ after the
    analysis cuts detailed in the text for the wino signal for
    $m_{1/2}=1.3$ and 1.6~TeV, with other parameters fixed as
    in Eq.~(\ref{eqn:mline}) is shown by the hollow histograms. The
    corresponding background distributions are shown by the filled
    histograms. The background histograms are not stacked.
  \label{fig:mT_2b}}
\end{center}
\end{figure}

Again, since only one boson is constructed in this channel, we show the
distributions of the transverse mass $m_T(bb,\eslt)$ in
Fig.~\ref{fig:mT_2b} for the two signal cases and for various SM
backgrounds. Not surprisingly, backgrounds from SM final states
involving $b$-quarks produced in association with vector bosons to give
sizeable $\eslt$ dominate: these include, $t\bar{t},Zb\bar{b},t\bar{t}Z$
and $Wb\bar{b}$ production.  The signals peak at $\sim 1400-1600$~GeV
depending on the wino mass, while the backgrounds are broader continua.
We see that in favourable cases the signal, though not large, is
comparable to the background in several bins and it appears that this
channel could make a contribution to the significance when the channels
are combined.

\subsection{$\ell^\pm \ell^\pm +\eslt$ channel}

This channel is designed to catch very characteristic events with two
same sign dileptons coming from the leptonic decays of same sign $W$
bosons produced from the decays of pair-produced
winos\cite{Baer:2013yha}. This signal-- which is characteristic of light
higgsino models-- has low rates but is interesting because it also has
very low backgrounds from SM processes. In Ref.\cite{Baer:2017gzf}, it
was shown that the discovery reach of the HL-LHC, via this single
channel, extended to a wino mass of $\sim 860$~GeV. Here, since we are
exploring the wino reach that might be possible by combining several
channels,  we have reanalysed the same-sign dilepton channel exploring
harder cuts that might allow us to go out further in the wino  mass at
the HL-LHC. Upon exploring various distributions, we require the additional
 cuts,
\begin{itemize} 

\item $|\eta(\ell)| < 2$;
\item $\eslt>350$~GeV;
\item $\Delta\phi (\ell\ell,\eslt) > \pi/3$ where $\Delta\phi$ is the
  transverse plane opening angle between the $\vec{p}_T(\ell\ell)$ and
  $\vec{\eslt}$;
\item  $\Delta\phi(\ell_1,\eslt)>\pi/4$ and $\Delta\phi(\ell_2,\eslt)>\pi/4$. 
\end{itemize}
The $\eslt$ cut allows us to probe the signal from TeV scale winos and
the angular cuts require that the $\eslt$ vector is well-separatd from
the leptons. 

\begin{figure}[htb!]
\begin{center}
  \includegraphics[height=0.4\textheight]{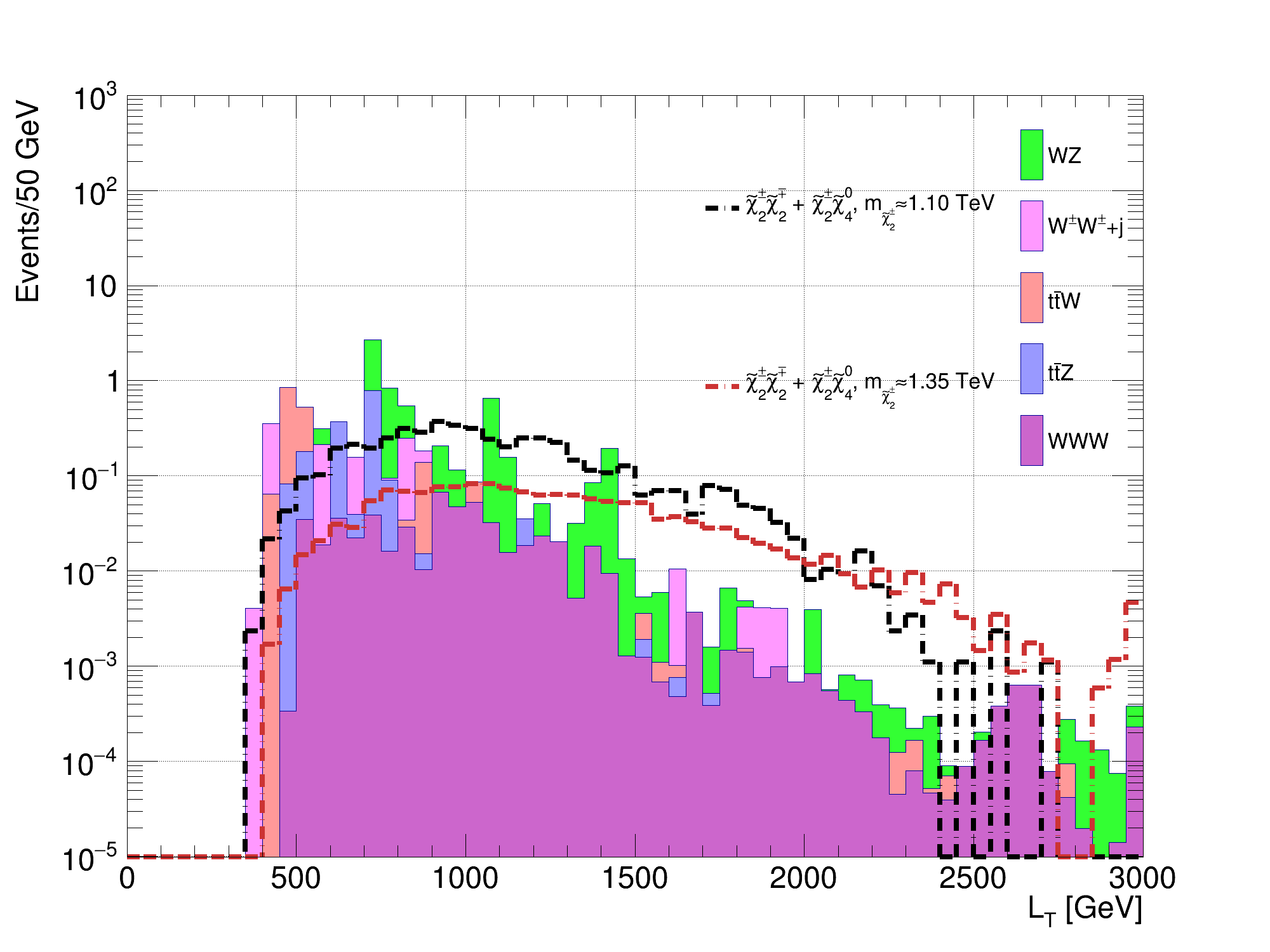}
  \caption{The distribution of $L_T\equiv
    |p_T(\ell_1)|+|p_T(\ell_2)|+|\eslt|$ after the analysis cuts
    detailed in the text for the wino signal for $m_{1/2}=1.3$
    and 1.6~TeV, with other parameters fixed as in Eq.~(\ref{eqn:mline})
    is shown by the hollow histograms. The corresponding background
    distributions are shown by the filled histograms. The background
    histograms are not stacked.
  \label{fig:LTSS}}
\end{center}
\end{figure}

In Fig.~\ref{fig:LTSS} we show the distribution of $L_T\equiv
|p_T(\ell_1)|+|p_T(\ell_2)|+|\eslt|$ for the two signal cases and for
dominant SM backgrounds after the analysis just described. We have
checked that $t\bar{t}$ production (not shown) makes a subdominant
contribution to the signal. We see that though the signal rate is very
small, the signal stands out above SM backgrounds for
$L_T>$~800-1000~GeV. We emphasize that this channel is characteristic of
models with light higgsinos whose decay products are essentially
invisible and would be absent in models where winos decayed to binos and
higgino states are decoupled. We also stress that the signal has
essentially no jet activity other than that from QCD radiation, and so
should be readily distinguishable from the same-sign dilepton production
via gluino\cite{Barnett:1987kn,Baer:1989hr,Baer:1991xs,Barnett:1993ea}
or squark\cite{Baer:1995va} pair production .

\section{Reach of HL-LHC for EWinos in natural SUSY}
\label{sec:reach}

Now that we have settled on our strategy to probe winos via the eight
channels discussed in Sec.~\ref{sec:channels}, it is possible to obtain
the LHC discovery sensitivity should there be an excess of events above SM
backgrounds, or the corresponding exclusion limit if no such excess is
observed at the HL-LHC. We express this in terms of the largest value of
$m_{1/2}$ (or equivalently, the wino mass) that can be probed in the
HL-LHC run which is envisaged to accumulate an integrated luminosity of
3000~fb$^{-1}$.

For each of the first seven channels we examine the binned $m_{T2}$ or
the transverse mass distributions shown in
Fig.~\ref{fig:mT2_2l1J}-\ref{fig:mT_2b}.  For the same-sign dilepton +
$\eslt$ channel, we examined the $L_T$ distribution in
Fig.~\ref{fig:LTSS}. For exclusion of the wino signal, we assume that
the true distribution we would observe in an experiment would correspond
to a background only distribution. Upper limits on $m_{1/2}$ are then
evaluated using a modified frequentist $CL_S$ method \cite{Read_2002}
with the profile likelihood ratio as the test statistic.  The likelihood
is built as a product of Poissonian terms for each of the bins in the
distributions. A background systematic error is accounted for by
introducing an independent nuisance parameter for each bin of each
channel and the likelihood is modified by log-normal terms to account
for these nuisance parameters, with uncertainty that we take to be 25\%.
The largest value of $m_{1/2}$ (or equivalently, the largest value of
wino mass) that can be excluded at 95\%CL is the exclusion limit.  For
discovery, we assume that the distribution one would observe in an
experiment corresponds to signal-plus-background. We then test this
against the background only distribution for each value of $m_{1/2}$.
If the background only hypothesis can be rejected at at least the
5$\sigma$ level, we deem that the HL-LHC would discover winos with a
mass corresponding to that choice of $m_{1/2}$.  For both the exclusion
and discovery limits, we use the asymptotic expansion for obtaining the
median significance \cite{Cowan_2011}.\footnote{We have checked that for
  every channel that we study there are at least ten (frequently
  significantly more) background events in the ``sensitive regions'' of
  the histograms in Fig.~\ref{fig:mT2_2l1J}-\ref{fig:LTSS}. This is
  large enough to justify the use of asymptotic formulae since for
  discovery (exclusion) we are concerned with fluctuations of the
  background (signal plus background).}

To warm up, we begin by considering the relative importance of the eight
channels introduced in Sec.~\ref{sec:channels} to the wino reach at the
HL-LHC. The presence of a signal will distort the $m_{T2}$,
$M_T$ or $L_T$ distributions illustrated in Sec.~\ref{sec:channels}. The
magnitude of this distortion and its statistical significance depends on
both the number of signal as well as the number of background events in
optimally chosen bins and the probability of the background fluctuating
to the level of the signal can be translated into the number of
``standard deviations''. Our results of this exercise are shown in
Table~\ref{tab:signif} for the case with $m_{1/2}=1.3$~TeV along our
model line, with statistical errors only as well as with an assumed 25\%
systematic error on the background: in this choice, we are
guided by Fig.~9 of the ATLAS study Ref.\cite{ATLAS:2021yqv}.
\begin{table}[h!]
\centering
\begin{tabular}{ccc}
\hline
Signal channel & Significance (0\% systematic) & Significance (25\% systematic) \\
\hline
$Z(\to \ell^+\ell^-) B+\eslt$ & 2.1 & 2.2\\
$h/Z(\to bb)B+\eslt$ & 2.6 & 2.4\\
$BB+\eslt$ & 1.60 &  0.4\\
$\ell h+\eslt$ & 1.6 & 1.5\\
$\ell B_{W/Z}+\eslt$ & 1.4 & 0.6\\
$Z(\ell^+\ell^-) +\eslt$ & 1.2 & 1.2\\
$bb+\eslt$ & 1.5 & 1.2\\
$\ell^\pm\ell^\pm+\eslt$ & 2.4 & 2.4\\
\hline
$combined$ & 5.3 & 4.7\\
\hline
\end{tabular}
\caption{Statistical significance of the signal for each of the eight
  different signal channels for the model-line case with $m_{1/2}=1.3$
  TeV at HL-LHC, assuming an integrated luminosity of 3000 fb$^{-1}$.}
\label{tab:signif}
\end{table}
We see that the $h/Z(\to bb)B+\eslt$ channel is the largest
contributor, closely followed by the same-sign dilepton +$\eslt$
channel. The other channels individually make smaller contributions
but combine to noticeably increase the significance.
We see from Fig.~\ref{fig:mT2_2l1J} and Fig.~\ref{fig:mT2_2b1J} that the
signal sticks out over the background (remember though the background
histograms are not stacked) but the higher signal rate leads to a
somewhat better significance in the second case. When there is no
systematic, we see that the $BB+\eslt$ channel remains competitive with
the remaining channels in spite of the fact that the signal in
Fig.~\ref{fig:mT2_2J} is buried under the background. As already noted,
this is due to the large signal rate expected in this channel. Once
systematic uncertainty is introduced, significance in channels that have
low $S/B$ ratio such as $BB+\eslt$ and $\ell B_{W/Z}+\eslt$ channels are
reduced significantly. Those with high $S/B$ ratio are more resilient
even if large systematic is present.  Although none of the channels
individually provide even $3\sigma$ evidence of a signal, we see from
the Table that by combining the eight channels one would attain a
$5\sigma$ significance if systematic uncertainties can be ignored.

\begin{figure}[htb!]
\begin{center}
  \includegraphics[height=0.35\textheight]{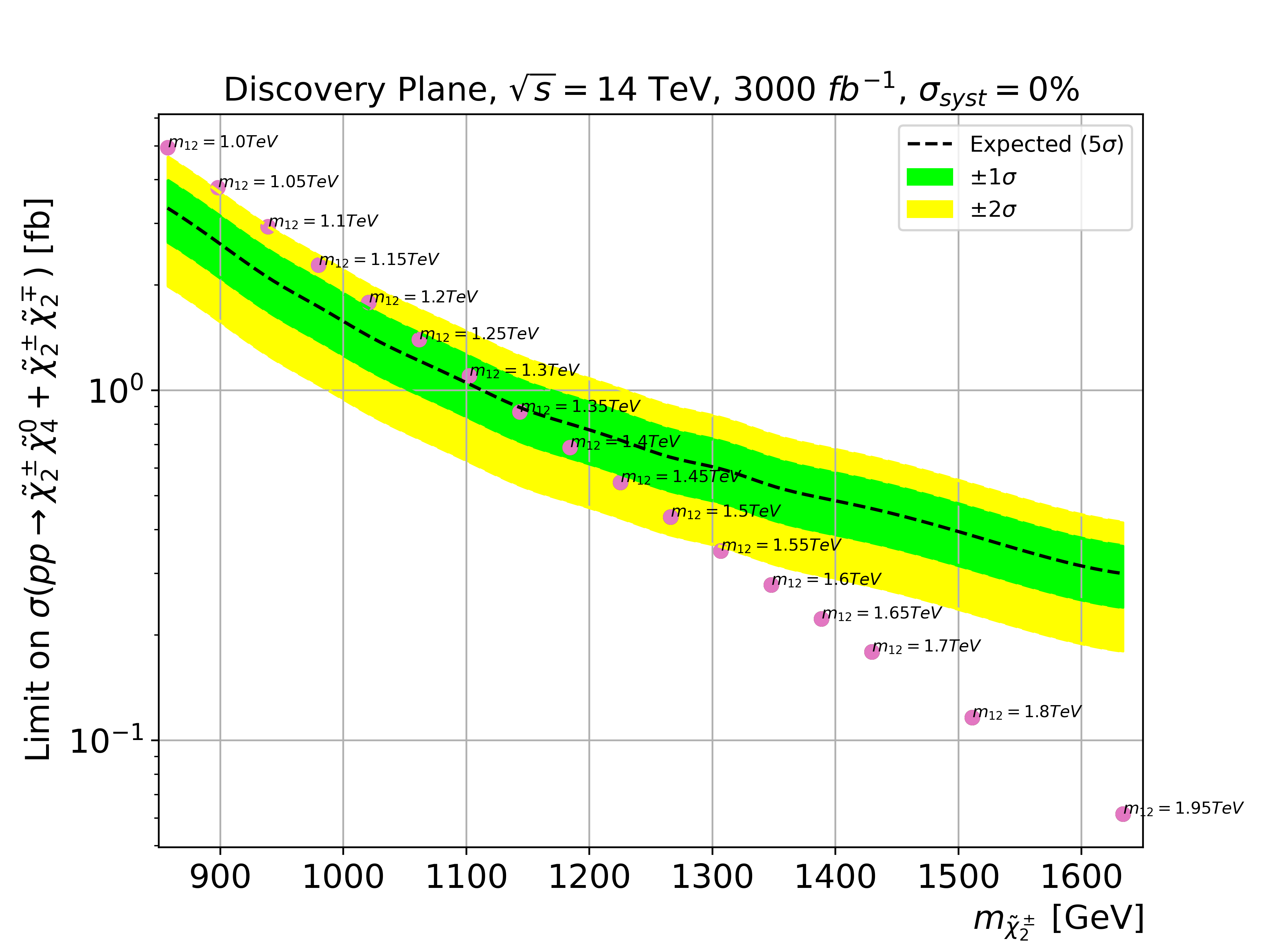}\\
  \includegraphics[height=0.35\textheight]{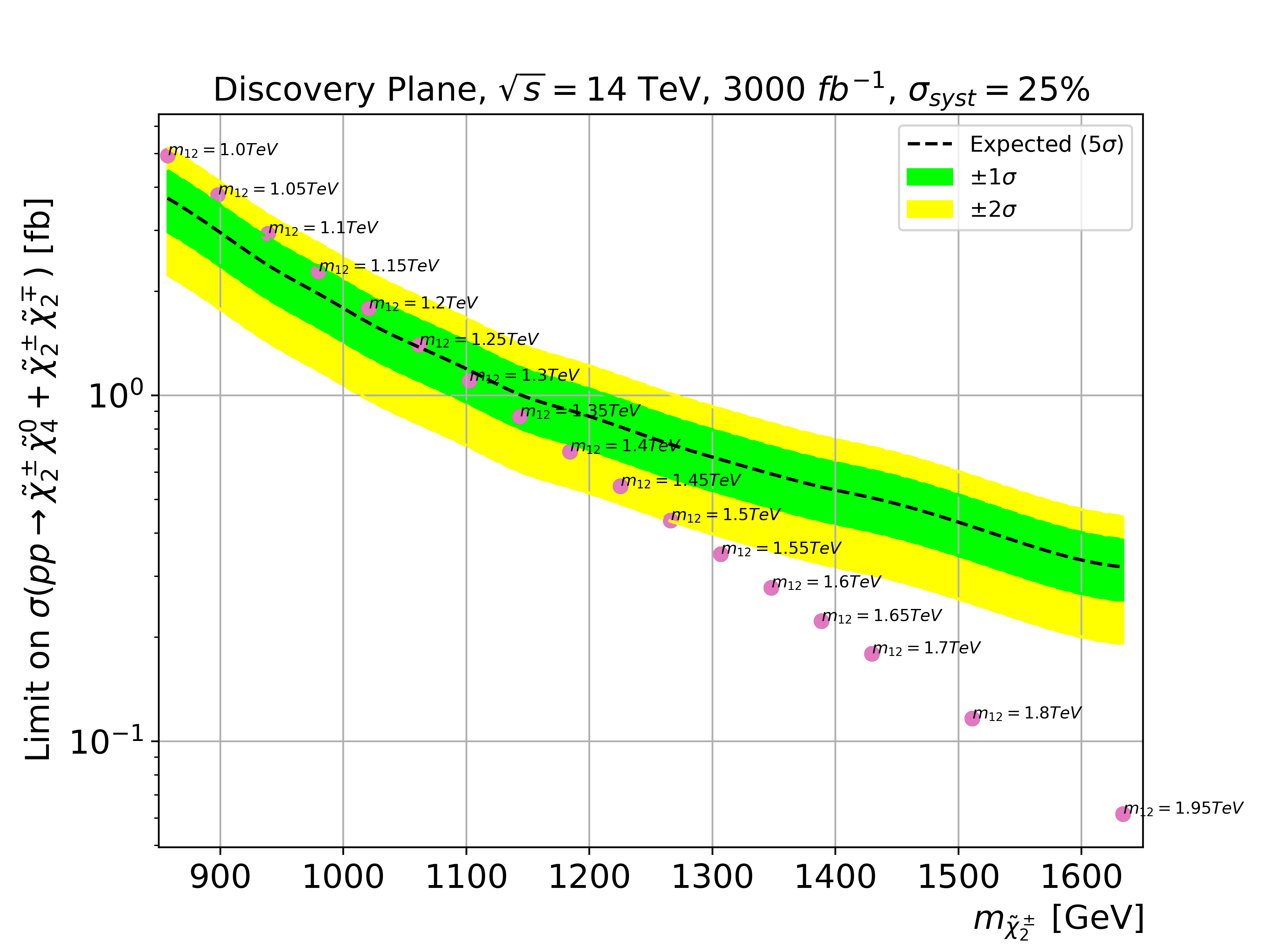}
  \caption{$5\sigma$ discovery reach of HL-LHC from wino pair production
    as a function of $m_{1/2}$, with other parameters as in
    Eq.~\ref{eqn:mline}, after combining the eight discovery channels
    detailed in the text. The upper frame shows the reach with
    statistical errors alone while the lower frame shows the reach
    assuming an additional 25\% systematic error common to all the
    channels.
  \label{fig:reach}}
\end{center}
\end{figure}
\begin{figure}[htb!]
\begin{center}
  \includegraphics[height=0.35\textheight]{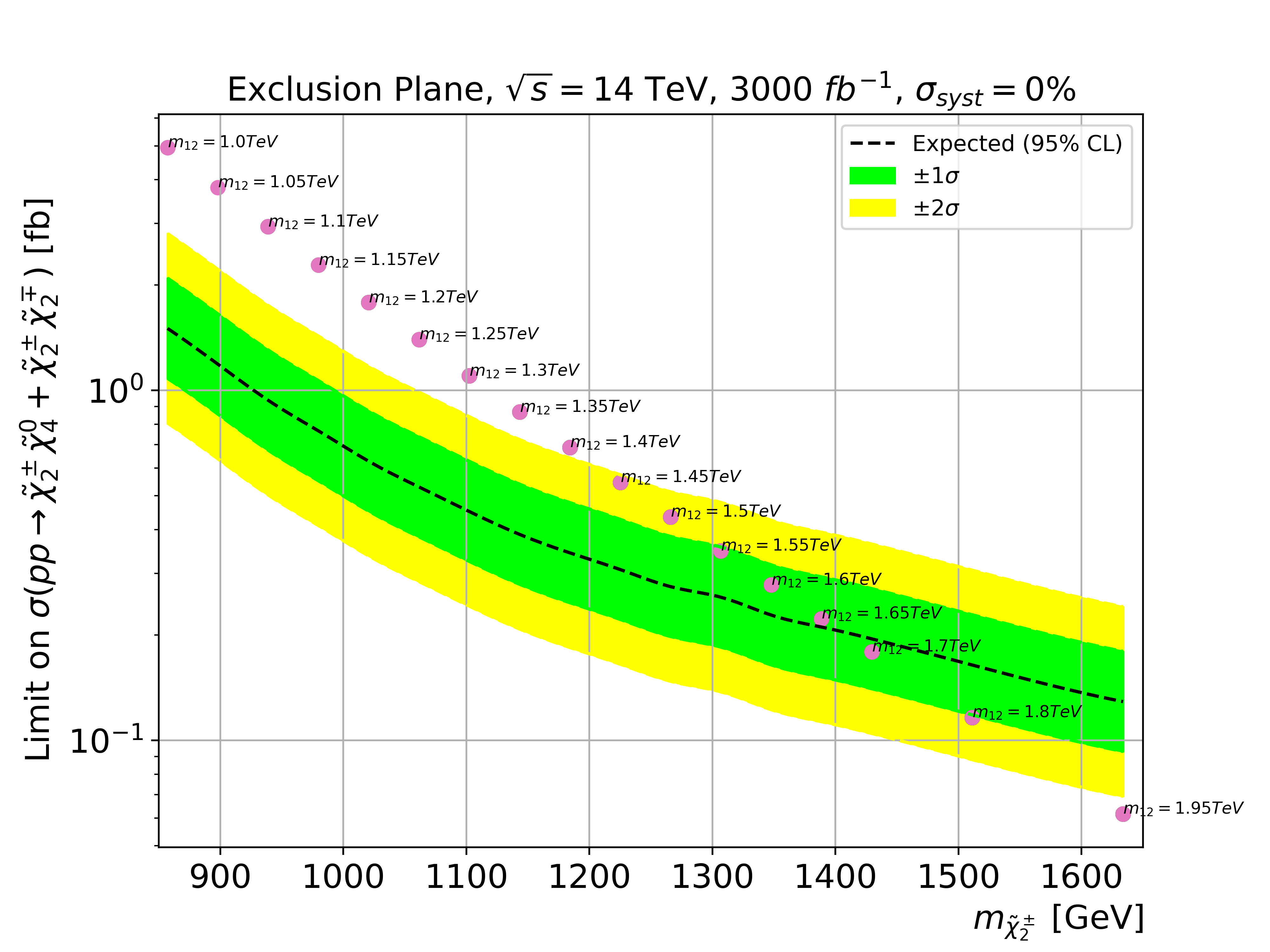}\\
  \includegraphics[height=0.35\textheight]{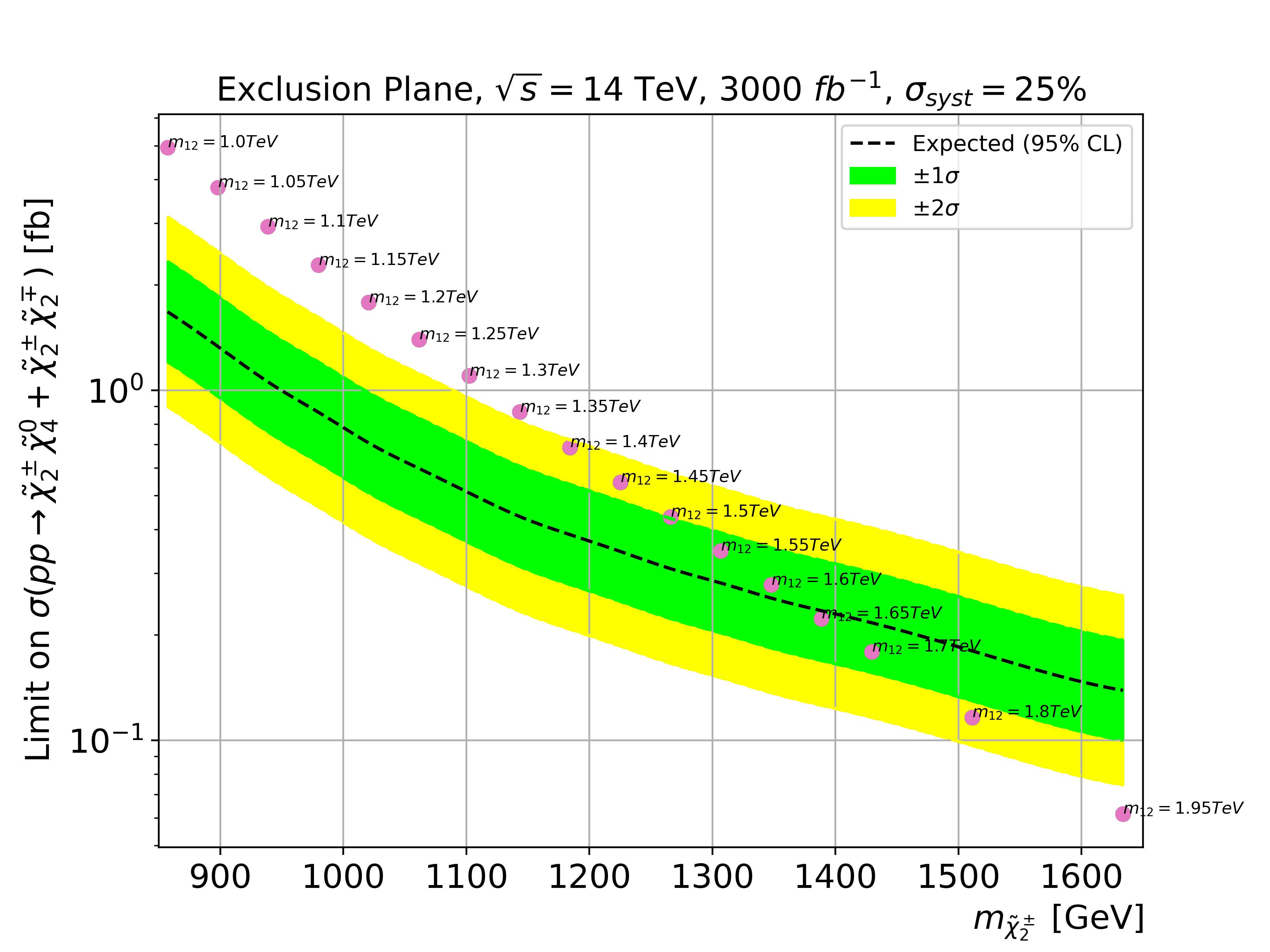}
  \caption{The 95\%CL exclusion limit of HL-LHC from wino pair
    production as a function of $m_{1/2}$, with other parameters as in
    Eq.~(\ref{eqn:mline}), after combining the eight discovery channels
    detailed in the text. The upper frame shows the reach with
    statistical errors alone while the lower frame shows the reach
    assuming an additional 25\% systematic error common to all the
    channels.
  \label{fig:exclusion}}
\end{center}
\end{figure}

The HL-LHC discovery reach and the 95\%CL exclusion level after
combining all the channels with statistical errors alone is shown in the
upper frames of Fig.~\ref{fig:reach} and Fig.~\ref{fig:exclusion},
respectively. The systematic error is almost certainly channel-dependent
and difficult to evaluate. To illustrate its impact, however, we show in
the bottom frames how these might be altered if we assume a common
systematic error of 25\% for all the channels. The minimum cross section
for discovery/exclusion is shown by the black dashed line; following
ATLAS and CMS, we denote by the green (yellow) bands how much this
discovery/exclusion line might move due to background fluctuations at
the $1\sigma$ ($2\sigma$) level. From Fig.~\ref{fig:reach}, we see from
the upper frame that the HL-LHC discovery limit for winos extends to a
wino mass of about 1.15~TeV. From the lower frame, we see that with the
assumed 25\% common systematic uncertainty, the discovery limit drops by
$\sim 50$~GeV. Turning to Fig.~\ref{fig:exclusion}, we project that
experiments at the HL-LHC would be sensitive to charged wino mass of
almost 1.4~TeV.  Although not shown, we have checked that even with a
100\% systematic uncertainty on the background, the exclusion contour
extends to $m_{\tw_2}= 1.3$~TeV.

Before closing this section, we point out that we have considered the
signal only from wino pair production in this paper even though the
cross section for charged-wino production in association with a bino,
shown in Fig.~\ref{fig:prod}({\it c}), 
is comparable to the cross section for the associated production of
charged  and neutral winos. The reason for this is that -- for the channels
with  hadronically decaying $W/Z/h$ bosons -- our analysis cuts work best for
boosted bosons. For the NUHM2 model with unified gaugino masses, the
bino mass is about half the wino mass, and the boson daughters from bino
decay have too small a boost to pass these cuts efficiently.

\section{Do wino signals leave an imprint of a light higgsino?}
\label{sec:higgsino}

Up to now, we have concentrated on the HL-LHC reach via signals from
wino production in natural SUSY models. In this section, we ask whether
it is possible to tell whether the winos are indeed decaying to
higgsinos (as they must in natural SUSY models because $|\mu|$ cannot be
much larger than the weak scale) should signals for winos appear at the
HL-LHC. We begin with a discussion of how the wino decay patterns would
be altered from those discussed in Sec.~\ref{sec:BF} if higgsinos are
very heavy and inaccesible via decays of winos; {\it i.e.} in models
where the LSP is dominantly bino-like. For simplicity of discussion, we
assume $|\mu|\gg M_2 \simeq 2M_1$, and continue to take sfermions to be
in the multi-TeV mass range. We also take $m_A = 2$~TeV so that the
additional Higgs bosons of the MSSM are also inaccessible via decays of
the winos. This leads to an electroweak-ino spectrum that was in vogue
in many early SUSY analyses performed within the so-called minimal
supergravity (mSUGRA) framework (see {\it e.g.} Ref.~\cite{Baer:2006rs})
with $\tz_1$ being bino-like, $\tw_1$ and $\tz_2$ being wino-like and
$\tw_2, \tz_3$ and $\tz_4$ being higgsino-like, with a sizeable gap
between $m_{\tw_1} \simeq m_{\tz_2}$ and $m_{\tz_1}$.

In this case, the charged wino dominantly decays via $\tw_1 \to
W\tz_1$, this being the only two-body decay accesible to it. There are
no two-body decays to a $Z$ or to $h$. This is in sharp contrast to
the situation shown in Fig.~\ref{fig:bfs} where we saw that the wino
state $\tw_2$ decayed into $W, Z$ and $h$ plus an quasi-invisible
higgsino with branching ratios of about 50\%, 25\% and 25\%,
respectively.

Turning our attention to the neutral wino state $\tz_2$, we note that
the decays $\tz_2 \to h\tz_1$ and $\tz_2\to Z\tz_1$ are both
kinematically accessible for TeV scale winos. Note, however, that the
$\tz_2-h-\tz_1$ coupling can only occur due to the higgsino component of
$\tz_1$ or $\tz_2$, and so is suppressed by a small mixing angle $\sim
m_Z/|\mu|$. In contrast, since the $Z$ couples to neutralinos only via
the higgsino components of both the neutralinos (gauge invariance
precludes a coupling of $Z$ to neutral gauginos), the coupling to $Z$ is
suppressed by two factors of the small mixing angle. As a result, in
models with $|\mu| \gg M_2 \simeq 2M_1$, the neutral wino decays almost
exclusively via $\tz_2\to \tz_1 h$; {\it i.e.} the branching ratio for
$\tz_2\to \tz_1 Z$ is {\em dynamically} suppressed.\footnote{We note
  that because of the gaugino mass unification assumption, the bino mass
  can never be neglected in the computation of the wino decay widths. As
  a result, neutral wino decays into the longitudinally polarized $Z$
  boson are not as enhanced as for the case of the higgsino LSP
  discussed in Ref.\cite{Baer:2017gzf} (see Eq.~(B.61b) of
  Ref.\cite{Baer:2006rs}), and the branching ratio for $\tz_2\to \tz_1
  Z$ decays remains small.  We have checked that $B(\tz_2\to \tz_1 Z)$,
  which also depends on mixing angles, is typically below $\sim 5$\%
  (10\%) for a wino mass of 1.7 (3)~TeV. This situation may be different
  in models without gaugino mass unification if the weak scale bino mass
  parameter is much smaller than the wino mass parameter. }

The upshot of this discussion is that in large $|\mu|$ models (at least
those with gaugino mass unification) {\em there cannot be a signal from
  wino production in those channels involving an identified high $p_T$
  $Z$ boson, whereas in models with light higgsinos, these signals must
  be present.} Of the eight channels introduced in
Sec.~\ref{sec:channels}, the channels, 1, $2^\prime$ and 6 clearly have
an identified $Z$ boson in them, where $2^\prime$ here denotes channel 2
with $80~{\rm GeV} < m_{bb} < 100$~GeV. Moreover, $W^\pm W^\pm+\eslt$
events required for events in channel 8 occur only if winos can decay
into higgsinos. A signal in channels, 1, $2^\prime$, 6 and 8 is thus a
clear indication of light higgsinos, assuming that the signal from the
eight channels originates in the production of wino states at the
HL-LHC.

\begin{figure}[htb!]
\begin{center}
  \includegraphics[height=0.4\textheight]{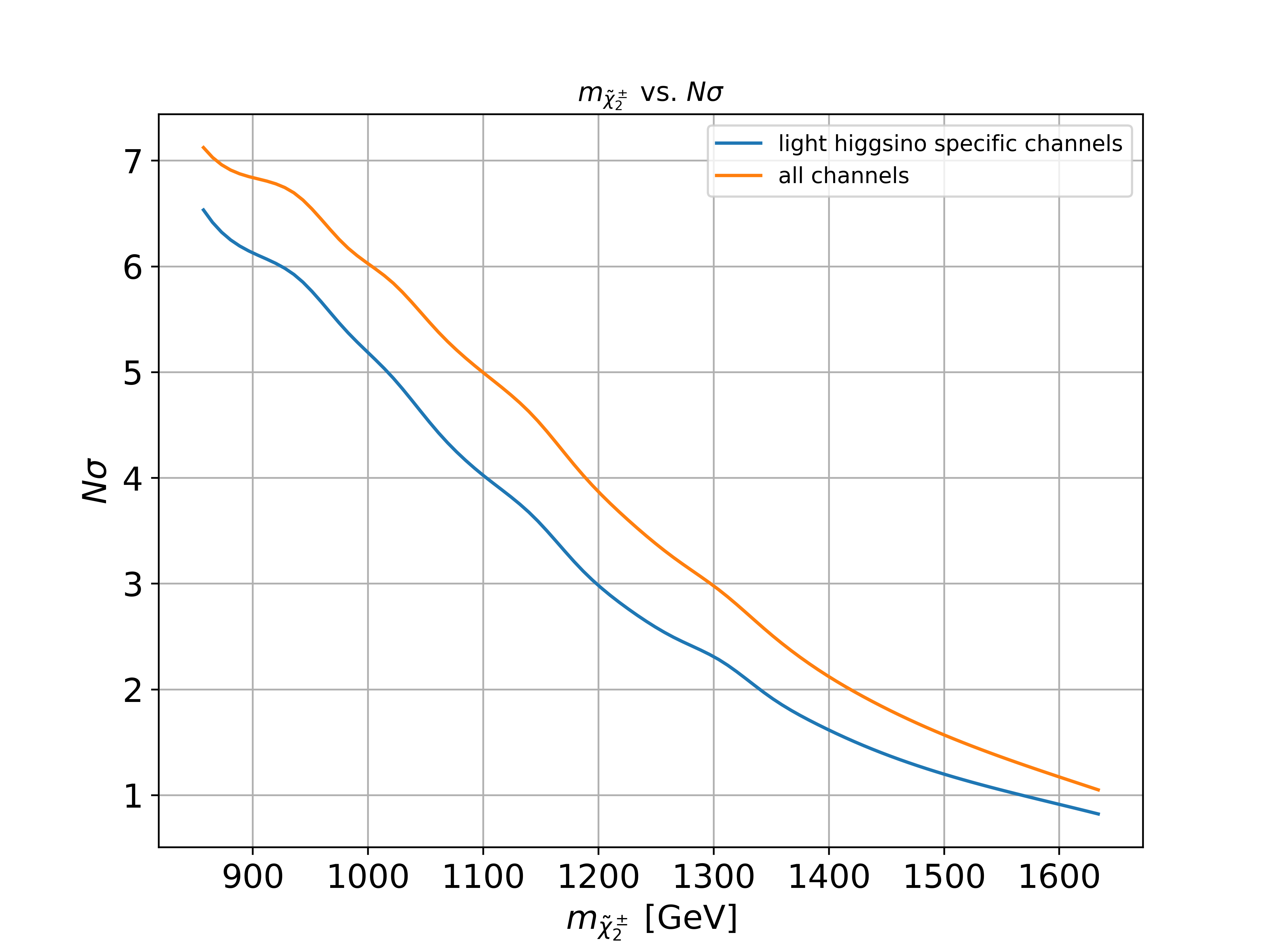}
  \caption{The statistical significance for wino pair production as a
    function of $m_{\tw_2}$, with other parameters as in
    Eq.~(\ref{eqn:mline}), after combining the eight discovery channels
    detailed in the text (upper curve), and combining only the light
    higgsino specific channels with an identified $Z$ boson (lower
    curve) including an additional 25\%
    systematic error common to all the channels.
  \label{fig:signif}}
\end{center}
\end{figure}

To quantify this, we show in Fig.~\ref{fig:signif} the statistical
significance of the signal above SM expectations as a function of the
wino mass. The upper curve shows the result obtained by combining all
eight channels, while the lower curve shows the result obtained
including the channels with a clearly identified $Z$ boson (channels 1,
$2^\prime$, 6 and 8). These are labeled as {\em light higgsino specific
  channels} on the figure. A systematic uncertainty of 25\% is included
in this figure. We see that while the discovery reach of the HL-LHC with
3000~fb$^{-1}$ extends to 1100~GeV, we interpret the lower curve as
indicative of about $3\sigma$ evidence for the existence of light
higgsinos out to a wino mass of 1200~GeV if we attribute the signal as
arising from winos of supersymmetry.

We mention here that events with same sign dileptons could also arise
from same sign wino production via $W^\pm W^\pm \to \tw_2\tw_2$
scattering: these events would be characterized by the presence of
energetic jets in the high $|\eta|$ region. Also, high $p_T$ $Z$ bosons
could arise in models with large $|\mu|$ from heavy higgsinos decaying
to the lighter inos, or even from gluino and squark cascades as pointed
out more than three decades ago \cite{Baer:1990in}. Note that in either
of these cases, higgsino states are necessary to get a $Z$ boson
daughter in the signal.\footnote{In principle, high $p_T$ $Z$ bosons can
  also occur via decays of heavy sfermions if there is large
  intra-generation mixing \cite{Baer:2006rs}, or if the super-GIM
  mechanism is not operative. Decays of heavy Higgs bosons, {\it e.g.}
  $A\to hZ$ could also lead to high $p_T$ $Z$ bosons in an event. Both
  sfermion events as well as heavy Higgs boson events involving $Z$
  would be readily distinguishable from wino events.} It is clear,
however, that the event rates from heavy higgsino cascades to winos
would be much smaller than the corresponding rates expected from wino
production in natural SUSY models. Gluino and squark events would be
distinguished by very different event topologies from wino events
studied in this paper. Events with high $p_T$ $Z$ bosons, together with
$\ell^\pm\ell^\pm+\eslt$ events with limited jet activity, will provide
indirect evidence for the existence of light higgsinos, should a wino
signal be found at the LHC.

Instead of comparing with the SM as we do in Fig.~\ref{fig:signif}, we
considered a comparison of the natural SUSY model with an NUHM2 model
with large $\mu$ so that the bino is the LSP. In this case, the gaugino
mass unification assumption reduces the wino-LSP mass gap from that in
the natural SUSY model with $\mu=250$~GeV. Setting the bino-LSP mass to
be 250~GeV, however, takes us close to the region currently
excluded by the LHC for winos up to about 1~TeV. A comparison with the
SM, keeping only the light higgsino specific channels, seems to be the
cleanest way to test for light higgsinos.

Another point of concern may be that this evidence dwindles for wino
masses not much above the current bounds from the LHC. While this is
true for the HL-LHC, we believe that an examination of these light
higgsino specific channels is nonetheless important because the
soft dilepton plus monojet signature which can produce
direct evidence for light higgsinos at the HL-LHC is very sensitive to
the mass gap between $\tz_1$ and $\tz_2$.
In contrast, the signal with
high $p_T$ $Z$ bosons is  insensitive to the size of the mass gap, and
could prove important at a future hadron collider with larger energy
and/or luminosity than the HL-LHC. At the very least, it provides
complimentary, albeit indirect, evidence for the existence of light
higgsinos. It goes almost without saying that electron-positron
colliders with sufficient centre-of-mass energy would provide the most
unambiguous evidence for light higgsinos \cite{Baer:2014yta,Moortgat-Pick:2015lbx,Baer:2016new,Lehtinen:2017vdt,Baer:2019gvu}.

\section{Summary and Concluding Remarks}
\label{sec:conclude}

In this paper we have continued our exploration of the reach of the
HL-LHC for superpartners in natural SUSY models, characterized by a value
of the electroweak fine-tuning measure $\Delta_{EW} < 30$. In these
scenarios, higgsinos are expected to be below $\sim 350$~GeV while
other superpartners, including top squarks, could well be in the
multi-TeV range, well beyond the reach of the LHC. While experiments at
the HL-LHC may indeed be able to directly probe higgsino signals via the
monojet plus soft dilepton channel, the prospects for discovery are
sensitive to the higgsino mass gap and their discovery is not guaranteed
for the entire range of SUSY parameters\cite{Baer:2020sgm}. Indeed, discovery of natural
SUSY is guaranteed only at future colliders, {\it e.g.} an
electron-positron collider with a centre-of-mass energy high enough to
produce higgsinos, or a high energy $pp$ collider with an energy in
excess of 27~TeV assuming it can accumulate an integrated luminosity of
15~ab$^{-1}$ \cite{Baer:2018hpb}. While it is not known whether either
an $e^+e^-$ or a hadron collider with the required energy will ever be
constructed, there are clear plans to increase the luminosity of the LHC
by an order of magnitude, and operate this machine to accumulate
3000~fb$^{-1}$ of integrated luminosity over a decade. In this paper, we
examine the discovery potential for winos of supersymmetry at the HL-LHC
within the natural SUSY framework.

For winos in the 1-2~TeV range, the cross sections for wino pair
production range between ${\cal O}(1)$~fb to ${\cal O}(0.01)$~fb, as
discussed in Sec.~\ref{sec:prod}. For heavy sfermions, these cross
sections are determined by the $SU(2)\times U(1)$ gauge interactions and
so are essentially fixed by the wino mass, independent of the SUSY
model. However, wino decay patterns -- and hence the signatures -- depend on
the nature of the LSP, and so are very different in natural SUSY from
the more well-studied case of a bino LSP. Assuming that matter sfermions and
the additional Higgs bosons are significantly heavier than the wino, in
natural SUSY heavy charged and neutral winos decay to on-shell $W, Z$
and $h$ bosons and an associated higgsino with branching fractions in
the ratio $\sim 2:1:1$: see Fig.~\ref{fig:bfs}. In contrast, in models
with a bino-like LSP, the charged wino essentially always decays to a
$W$ boson and an LSP, while the neutral wino decays to an $h$ boson and
an LSP. Since the higgsino mass gap is typically $4-15$~GeV in natural
SUSY models, the visible decay products of the daughter higgsinos are
too soft to be detected (without special effort) at hadron colliders,
and wino pair production is signalled by $VV$, $Vh$ and $hh$ plus
$\eslt$ events in natural SUSY models.

To facilitate our study of winos in natural SUSY models, we have
identified eight experimentally distinct channels via which it would be
possible to search for wino pair production in Sec.~\ref{sec:channels}.
These channels depend on how the high $p_T$ $W, Z$ and $h$ boson
daughters of the TeV scale winos decay. For each of these channels, we
identify cuts that enhance the signal relative to SM backgrounds from
$t\bar{t}$, $t\bar{t}V$, $t\bar{t}h$, $VV$, $hh$, $VVV$, $hhV$, $V+jets$
and $Vh$ parton level processes. For each channel, we then plot one of
the $m_T$, $m_{T2}$ or $L_T$ distributions for various signal cases and
for SM backgrounds: these are shown in
Fig.~\ref{fig:mT2_2l1J} -- Fig.~\ref{fig:LTSS}.

In Sec.~\ref{sec:reach} we use these distributions to map out the
$5\sigma$ discovery and the 95\% CL exclusion region for winos at the
HL-LHC obtained after combining the signal from all eight
channels. Table~\ref{tab:signif} shows the relative importance of each
of these channels.  Our final results for the HL-LHC reach and exclusion
are summarized in Fig.~\ref{fig:reach} and Fig.~\ref{fig:exclusion},
respectively. The upper frame shows the results with statistical errors
alone, while the lower frame shows how these are affected if we assume a
common systematic error of 25\% for each of the eight channels. While
experiments at the HL-LHC may be able to discover (exclude at 95\%CL)
wino masses up to 1.1~TeV (1.4~TeV), this is unfortunately a small part
of the range allowed by natural SUSY. This should not be surprising
because wino masses are relatively weakly constrained by naturalness
considerations alone, and their discovery at the energy of the LHC would
have to be somewhat fortuitous.

While the HL-LHC can probe only a small part of the natural SUSY
parameter space via wino searches, our discussion in
Sec.~\ref{sec:higgsino} shows that if a natural SUSY wino signal is seen
at the LHC, it could potentially also provide indications for the
existence of light higgsinos: see Fig.~\ref{fig:signif}. The point is
that there are several higgsino-specific channels (labeled 1,
$2^\prime$, 6 and 8 in the text)
where instead there would be only a tiny signal from wino production in any
model with a bino LSP. While the range of parameters where this may be
possible at the HL-LHC is small, we find it very interesting that the
presence of light higgsinos may reveal itself at a hadron collider via
signals from wino pair production alone. This could be critical if the
next accelerator facility is a higher energy/luminosity hadron collider
and the higgsino mass gap is too small for the monojet plus soft
dilepton signal from higgsino pair production to be observable. At the
very least, the signal from winos yields an indirect confirmation of the
existence of light higgsinos.

{\it Acknowledgements:} 

This material is based upon work supported by the U.S. Department of
Energy, Office of Science, Office of High Energy Physics under Award
Number DE-SC-0009956. VB gratefully acknowledges
support from the William F. Vilas estate.


\bibliography{EWinos}

\begin{thebibliography}{10}
\expandafter\ifx\csname url\endcsname\relax
  \def\url#1{\texttt{#1}}\fi
\expandafter\ifx\csname urlprefix\endcsname\relax\def\urlprefix{URL }\fi
\expandafter\ifx\csname href\endcsname\relax
  \def\href#1#2{#2} \def\path#1{#1}\fi

\bibitem{Witten:1981nf}
E.~Witten, {Dynamical Breaking of Supersymmetry}, Nucl. Phys. B 188 (1981) 513.
\newblock \href {https://doi.org/10.1016/0550-3213(81)90006-7}
  {\path{doi:10.1016/0550-3213(81)90006-7}}.

\bibitem{Dimopoulos:1981zb}
S.~Dimopoulos, H.~Georgi, {Softly Broken Supersymmetry and SU(5)}, Nucl. Phys.
  B 193 (1981) 150--162.
\newblock \href {https://doi.org/10.1016/0550-3213(81)90522-8}
  {\path{doi:10.1016/0550-3213(81)90522-8}}.

\bibitem{Sakai:1981gr}
N.~Sakai, {Naturalness in Supersymmetric Guts}, Z. Phys. C 11 (1981) 153.
\newblock \href {https://doi.org/10.1007/BF01573998}
  {\path{doi:10.1007/BF01573998}}.

\bibitem{Kaul:1981hi}
R.~K. Kaul, P.~Majumdar, {Cancellation of Quadratically Divergent Mass
  Corrections in Globally Supersymmetric Spontaneously Broken Gauge Theories},
  Nucl. Phys. B 199 (1982) 36.
\newblock \href {https://doi.org/10.1016/0550-3213(82)90565-X}
  {\path{doi:10.1016/0550-3213(82)90565-X}}.

\bibitem{ATLAS:2012yve}
G.~Aad, et~al., {Observation of a new particle in the search for the Standard
  Model Higgs boson with the ATLAS detector at the LHC}, Phys. Lett. B 716
  (2012) 1--29.
\newblock \href {http://arxiv.org/abs/1207.7214} {\path{arXiv:1207.7214}},
  \href {https://doi.org/10.1016/j.physletb.2012.08.020}
  {\path{doi:10.1016/j.physletb.2012.08.020}}.

\bibitem{CMS:2012qbp}
S.~Chatrchyan, et~al., {Observation of a New Boson at a Mass of 125 GeV with
  the CMS Experiment at the LHC}, Phys. Lett. B 716 (2012) 30--61.
\newblock \href {http://arxiv.org/abs/1207.7235} {\path{arXiv:1207.7235}},
  \href {https://doi.org/10.1016/j.physletb.2012.08.021}
  {\path{doi:10.1016/j.physletb.2012.08.021}}.

\bibitem{Vissani:1997ys}
F.~Vissani, {Do experiments suggest a hierarchy problem?}, Phys. Rev. D 57
  (1998) 7027--7030.
\newblock \href {http://arxiv.org/abs/hep-ph/9709409}
  {\path{arXiv:hep-ph/9709409}}, \href
  {https://doi.org/10.1103/PhysRevD.57.7027}
  {\path{doi:10.1103/PhysRevD.57.7027}}.

\bibitem{ATLAS:2021twp}
G.~Aad, et~al., {Search for squarks and gluinos in final states with one
  isolated lepton, jets, and missing transverse momentum at $\sqrt{s}=13$~ with
  the ATLAS detector}, Eur. Phys. J. C 81~(7) (2021) 600.
\newblock \href {http://arxiv.org/abs/2101.01629} {\path{arXiv:2101.01629}},
  \href {https://doi.org/10.1140/epjc/s10052-021-09748-8}
  {\path{doi:10.1140/epjc/s10052-021-09748-8}}.

\bibitem{ATLAS:2022ihe}
G.~Aad, et~al., {Search for supersymmetry in final states with missing
  transverse momentum and three or more b-jets in 139 fb$^{-1}$ of
  proton\textendash{}proton collisions at $\sqrt{s} = 13$~TeV with the ATLAS
  detector}, Eur. Phys. J. C 83~(7) (2023) 561.
\newblock \href {http://arxiv.org/abs/2211.08028} {\path{arXiv:2211.08028}},
  \href {https://doi.org/10.1140/epjc/s10052-023-11543-6}
  {\path{doi:10.1140/epjc/s10052-023-11543-6}}.

\bibitem{CMS:2019ybf}
A.~M. Sirunyan, et~al., {Searches for physics beyond the standard model with
  the $M_\mathrm{T2}$ variable in hadronic final states with and without
  disappearing tracks in proton-proton collisions at $\sqrt{s}=$ 13 TeV}, Eur.
  Phys. J. C 80~(1) (2020) 3.
\newblock \href {http://arxiv.org/abs/1909.03460} {\path{arXiv:1909.03460}},
  \href {https://doi.org/10.1140/epjc/s10052-019-7493-x}
  {\path{doi:10.1140/epjc/s10052-019-7493-x}}.

\bibitem{CMS:2021beq}
A.~M. Sirunyan, et~al., {Search for top squark production in fully-hadronic
  final states in proton-proton collisions at $\sqrt{s} =$ 13 TeV}, Phys. Rev.
  D 104~(5) (2021) 052001.
\newblock \href {http://arxiv.org/abs/2103.01290} {\path{arXiv:2103.01290}},
  \href {https://doi.org/10.1103/PhysRevD.104.052001}
  {\path{doi:10.1103/PhysRevD.104.052001}}.

\bibitem{CMS:2022sfi}
A.~Tumasyan, et~al., {Search for electroweak production of charginos and
  neutralinos at s=13TeV in final states containing hadronic decays of WW, WZ,
  or WH and missing transverse momentum}, Phys. Lett. B 842 (2023) 137460.
\newblock \href {http://arxiv.org/abs/2205.09597} {\path{arXiv:2205.09597}},
  \href {https://doi.org/10.1016/j.physletb.2022.137460}
  {\path{doi:10.1016/j.physletb.2022.137460}}.

\bibitem{CMS:2021few}
A.~Tumasyan, et~al., {Search for chargino-neutralino production in events with
  Higgs and W bosons using 137 fb$^{-1}$ of proton-proton collisions at
  $\sqrt{s} = 13$ TeV}, JHEP 10 (2021) 045.
\newblock \href {http://arxiv.org/abs/2107.12553} {\path{arXiv:2107.12553}},
  \href {https://doi.org/10.1007/JHEP10(2021)045}
  {\path{doi:10.1007/JHEP10(2021)045}}.

\bibitem{CMS:2021cox}
A.~Tumasyan, et~al., {Search for electroweak production of charginos and
  neutralinos in proton-proton collisions at $ \sqrt{s} $ = 13 TeV}, JHEP 04
  (2022) 147.
\newblock \href {http://arxiv.org/abs/2106.14246} {\path{arXiv:2106.14246}},
  \href {https://doi.org/10.1007/JHEP04(2022)147}
  {\path{doi:10.1007/JHEP04(2022)147}}.

\bibitem{CMS:2023qhl}
{Combined search for electroweak production of winos, binos, higgsinos, and
  sleptons in proton-proton collisions at $\sqrt{s}= 13$ TeV,
  CMS-PAS-SUS-21-008 (2023).} (2023).

\bibitem{ATLAS:2020pgy}
G.~Aad, et~al., {Search for direct production of electroweakinos in final
  states with one lepton, missing transverse momentum and a Higgs boson
  decaying into two $b$-jets in $pp$ collisions at $\sqrt{s}=13$ TeV with the
  ATLAS detector}, Eur. Phys. J. C 80~(8) (2020) 691.
\newblock \href {http://arxiv.org/abs/1909.09226} {\path{arXiv:1909.09226}},
  \href {https://doi.org/10.1140/epjc/s10052-020-8050-3}
  {\path{doi:10.1140/epjc/s10052-020-8050-3}}.

\bibitem{ATLAS:2021yqv}
G.~Aad, et~al., {Search for charginos and neutralinos in final states with two
  boosted hadronically decaying bosons and missing transverse momentum in $pp$
  collisions at $\sqrt {s} = 13$ TeV with the ATLAS detector}, Phys. Rev. D
  104~(11) (2021) 112010.
\newblock \href {http://arxiv.org/abs/2108.07586} {\path{arXiv:2108.07586}},
  \href {https://doi.org/10.1103/PhysRevD.104.112010}
  {\path{doi:10.1103/PhysRevD.104.112010}}.

\bibitem{ATLAS:2022hbt}
G.~Aad, et~al., {Search for direct pair production of sleptons and charginos
  decaying to two leptons and neutralinos with mass splittings near the W-boson
  mass in $ \sqrt{s} = 13$ TeV pp collisions with the ATLAS detector}, JHEP 06
  (2023) 031.
\newblock \href {http://arxiv.org/abs/2209.13935} {\path{arXiv:2209.13935}},
  \href {https://doi.org/10.1007/JHEP06(2023)031}
  {\path{doi:10.1007/JHEP06(2023)031}}.

\bibitem{ATLAS:2022zwa}
G.~Aad, et~al., {Searches for new phenomena in events with two leptons, jets,
  and missing transverse momentum in 139~fb$^{-1}$ of $\sqrt{s}=13$~TeV $pp$
  collisions with the ATLAS detector}, Eur. Phys. J. C 83~(6) (2023) 515.
\newblock \href {http://arxiv.org/abs/2204.13072} {\path{arXiv:2204.13072}},
  \href {https://doi.org/10.1140/epjc/s10052-023-11434-w}
  {\path{doi:10.1140/epjc/s10052-023-11434-w}}.

\bibitem{Ellis:1986yg}
J.~R. Ellis, K.~Enqvist, D.~V. Nanopoulos, F.~Zwirner, {Observables in
  Low-Energy Superstring Models}, Mod. Phys. Lett. A 1 (1986) 57.
\newblock \href {https://doi.org/10.1142/S0217732386000105}
  {\path{doi:10.1142/S0217732386000105}}.

\bibitem{Barbieri:1987fn}
R.~Barbieri, G.~F. Giudice, {Upper Bounds on Supersymmetric Particle Masses},
  Nucl. Phys. B 306 (1988) 63--76.
\newblock \href {https://doi.org/10.1016/0550-3213(88)90171-X}
  {\path{doi:10.1016/0550-3213(88)90171-X}}.

\bibitem{Dimopoulos:1995mi}
S.~Dimopoulos, G.~F. Giudice, {Naturalness constraints in supersymmetric
  theories with nonuniversal soft terms}, Phys. Lett. B 357 (1995) 573--578.
\newblock \href {http://arxiv.org/abs/hep-ph/9507282}
  {\path{arXiv:hep-ph/9507282}}, \href
  {https://doi.org/10.1016/0370-2693(95)00961-J}
  {\path{doi:10.1016/0370-2693(95)00961-J}}.

\bibitem{Feng:2013pwa}
J.~L. Feng, {Naturalness and the Status of Supersymmetry}, Ann. Rev. Nucl.
  Part. Sci. 63 (2013) 351--382.
\newblock \href {http://arxiv.org/abs/1302.6587} {\path{arXiv:1302.6587}},
  \href {https://doi.org/10.1146/annurev-nucl-102010-130447}
  {\path{doi:10.1146/annurev-nucl-102010-130447}}.

\bibitem{Baer:2023cvi}
H.~Baer, V.~Barger, D.~Martinez, S.~Salam, {Practical naturalness and its
  implications for weak scale supersymmetry}, Phys. Rev. D 108~(3) (2023)
  035050.
\newblock \href {http://arxiv.org/abs/2305.16125} {\path{arXiv:2305.16125}},
  \href {https://doi.org/10.1103/PhysRevD.108.035050}
  {\path{doi:10.1103/PhysRevD.108.035050}}.

\bibitem{Baer:2013gva}
H.~Baer, V.~Barger, D.~Mickelson, {How conventional measures overestimate
  electroweak fine-tuning in supersymmetric theory}, Phys. Rev. D 88~(9) (2013)
  095013.
\newblock \href {http://arxiv.org/abs/1309.2984} {\path{arXiv:1309.2984}},
  \href {https://doi.org/10.1103/PhysRevD.88.095013}
  {\path{doi:10.1103/PhysRevD.88.095013}}.

\bibitem{Mustafayev:2014lqa}
A.~Mustafayev, X.~Tata, {Supersymmetry, Naturalness, and Light Higgsinos},
  Indian J. Phys. 88 (2014) 991--1004.
\newblock \href {http://arxiv.org/abs/1404.1386} {\path{arXiv:1404.1386}},
  \href {https://doi.org/10.1007/s12648-014-0504-8}
  {\path{doi:10.1007/s12648-014-0504-8}}.

\bibitem{Baer:2012cf}
H.~Baer, V.~Barger, P.~Huang, D.~Mickelson, A.~Mustafayev, X.~Tata, {Radiative
  natural supersymmetry: Reconciling electroweak fine-tuning and the Higgs
  boson mass}, Phys. Rev. D 87~(11) (2013) 115028.
\newblock \href {http://arxiv.org/abs/1212.2655} {\path{arXiv:1212.2655}},
  \href {https://doi.org/10.1103/PhysRevD.87.115028}
  {\path{doi:10.1103/PhysRevD.87.115028}}.

\bibitem{Baer:2021tta}
H.~Baer, V.~Barger, D.~Martinez, {Comparison of SUSY spectra generators for
  natural SUSY and string landscape predictions}, Eur. Phys. J. C 82~(2) (2022)
  172.
\newblock \href {http://arxiv.org/abs/2111.03096} {\path{arXiv:2111.03096}},
  \href {https://doi.org/10.1140/epjc/s10052-022-10141-2}
  {\path{doi:10.1140/epjc/s10052-022-10141-2}}.

\bibitem{Dedes:2002dy}
A.~Dedes, P.~Slavich, {Two loop corrections to radiative electroweak symmetry
  breaking in the MSSM}, Nucl. Phys. B 657 (2003) 333--354.
\newblock \href {http://arxiv.org/abs/hep-ph/0212132}
  {\path{arXiv:hep-ph/0212132}}, \href
  {https://doi.org/10.1016/S0550-3213(03)00173-1}
  {\path{doi:10.1016/S0550-3213(03)00173-1}}.

\bibitem{Bae:2019dgg}
K.~J. Bae, H.~Baer, V.~Barger, D.~Sengupta, {Revisiting the SUSY $\mu$ problem
  and its solutions in the LHC era}, Phys. Rev. D 99~(11) (2019) 115027.
\newblock \href {http://arxiv.org/abs/1902.10748} {\path{arXiv:1902.10748}},
  \href {https://doi.org/10.1103/PhysRevD.99.115027}
  {\path{doi:10.1103/PhysRevD.99.115027}}.

\bibitem{Baer:2012up}
H.~Baer, V.~Barger, P.~Huang, A.~Mustafayev, X.~Tata, {Radiative natural SUSY
  with a 125 GeV Higgs boson}, Phys. Rev. Lett. 109 (2012) 161802.
\newblock \href {http://arxiv.org/abs/1207.3343} {\path{arXiv:1207.3343}},
  \href {https://doi.org/10.1103/PhysRevLett.109.161802}
  {\path{doi:10.1103/PhysRevLett.109.161802}}.

\bibitem{Baer:2019cae}
H.~Baer, V.~Barger, S.~Salam, {Naturalness versus stringy naturalness (with
  implications for collider and dark matter searches}, Phys. Rev. Research. 1
  (2019) 023001.
\newblock \href {http://arxiv.org/abs/1906.07741} {\path{arXiv:1906.07741}},
  \href {https://doi.org/10.1103/PhysRevResearch.1.023001}
  {\path{doi:10.1103/PhysRevResearch.1.023001}}.

\bibitem{Agrawal:1998xa}
V.~Agrawal, S.~M. Barr, J.~F. Donoghue, D.~Seckel, {Anthropic considerations in
  multiple domain theories and the scale of electroweak symmetry breaking},
  Phys. Rev. Lett. 80 (1998) 1822--1825.
\newblock \href {http://arxiv.org/abs/hep-ph/9801253}
  {\path{arXiv:hep-ph/9801253}}, \href
  {https://doi.org/10.1103/PhysRevLett.80.1822}
  {\path{doi:10.1103/PhysRevLett.80.1822}}.

\bibitem{Baer:2020kwz}
H.~Baer, V.~Barger, S.~Salam, D.~Sengupta, K.~Sinha, {Status of weak scale
  supersymmetry after LHC Run 2 and ton-scale noble liquid WIMP searches}, Eur.
  Phys. J. ST 229~(21) (2020) 3085--3141.
\newblock \href {http://arxiv.org/abs/2002.03013} {\path{arXiv:2002.03013}},
  \href {https://doi.org/10.1140/epjst/e2020-000020-x}
  {\path{doi:10.1140/epjst/e2020-000020-x}}.

\bibitem{Baer:2015rja}
H.~Baer, V.~Barger, M.~Savoy, {Upper bounds on sparticle masses from
  naturalness or how to disprove weak scale supersymmetry}, Phys. Rev. D 93~(3)
  (2016) 035016.
\newblock \href {http://arxiv.org/abs/1509.02929} {\path{arXiv:1509.02929}},
  \href {https://doi.org/10.1103/PhysRevD.93.035016}
  {\path{doi:10.1103/PhysRevD.93.035016}}.

\bibitem{Baer:2016wkz}
H.~Baer, V.~Barger, J.~S. Gainer, P.~Huang, M.~Savoy, D.~Sengupta, X.~Tata,
  {Gluino reach and mass extraction at the LHC in radiatively-driven natural
  SUSY}, Eur. Phys. J. C 77~(7) (2017) 499.
\newblock \href {http://arxiv.org/abs/1612.00795} {\path{arXiv:1612.00795}},
  \href {https://doi.org/10.1140/epjc/s10052-017-5067-3}
  {\path{doi:10.1140/epjc/s10052-017-5067-3}}.

\bibitem{CidVidal:2018eel}
X.~Cid~Vidal, et~al., {Report from Working Group 3}: {Beyond the Standard Model
  physics at the HL-LHC and HE-LHC}, CERN Yellow Rep. Monogr. 7 (2019)
  585--865.
\newblock \href {http://arxiv.org/abs/1812.07831} {\path{arXiv:1812.07831}},
  \href {https://doi.org/10.23731/CYRM-2019-007.585}
  {\path{doi:10.23731/CYRM-2019-007.585}}.

\bibitem{Baer:2023uwo}
H.~Baer, V.~Barger, J.~Dutta, D.~Sengupta, K.~Zhang, {Top squarks from the
  landscape at high luminosity LHC} (7 2023).
\newblock \href {http://arxiv.org/abs/2307.08067} {\path{arXiv:2307.08067}}.

\bibitem{Baer:2011ec}
H.~Baer, V.~Barger, P.~Huang, {Hidden SUSY at the LHC: the light higgsino-world
  scenario and the role of a lepton collider}, JHEP 11 (2011) 031.
\newblock \href {http://arxiv.org/abs/1107.5581} {\path{arXiv:1107.5581}},
  \href {https://doi.org/10.1007/JHEP11(2011)031}
  {\path{doi:10.1007/JHEP11(2011)031}}.

\bibitem{Han:2014kaa}
Z.~Han, G.~D. Kribs, A.~Martin, A.~Menon, {Hunting quasidegenerate Higgsinos},
  Phys. Rev. D 89~(7) (2014) 075007.
\newblock \href {http://arxiv.org/abs/1401.1235} {\path{arXiv:1401.1235}},
  \href {https://doi.org/10.1103/PhysRevD.89.075007}
  {\path{doi:10.1103/PhysRevD.89.075007}}.

\bibitem{Baer:2014kya}
H.~Baer, A.~Mustafayev, X.~Tata, {Monojet plus soft dilepton signal from light
  higgsino pair production at LHC14}, Phys. Rev. D 90~(11) (2014) 115007.
\newblock \href {http://arxiv.org/abs/1409.7058} {\path{arXiv:1409.7058}},
  \href {https://doi.org/10.1103/PhysRevD.90.115007}
  {\path{doi:10.1103/PhysRevD.90.115007}}.

\bibitem{Han:2015lma}
C.~Han, D.~Kim, S.~Munir, M.~Park, {Accessing the core of naturalness, nearly
  degenerate higgsinos, at the LHC}, JHEP 04 (2015) 132.
\newblock \href {http://arxiv.org/abs/1502.03734} {\path{arXiv:1502.03734}},
  \href {https://doi.org/10.1007/JHEP04(2015)132}
  {\path{doi:10.1007/JHEP04(2015)132}}.

\bibitem{ATLAS:2019lng}
G.~Aad, et~al., {Searches for electroweak production of supersymmetric
  particles with compressed mass spectra in $\sqrt{s}=$ 13 TeV $pp$ collisions
  with the ATLAS detector}, Phys. Rev. D 101~(5) (2020) 052005.
\newblock \href {http://arxiv.org/abs/1911.12606} {\path{arXiv:1911.12606}},
  \href {https://doi.org/10.1103/PhysRevD.101.052005}
  {\path{doi:10.1103/PhysRevD.101.052005}}.

\bibitem{CMS:2021edw}
A.~Tumasyan, et~al., {Search for supersymmetry in final states with two or
  three soft leptons and missing transverse momentum in proton-proton
  collisions at $ \sqrt{s} = 13$ TeV}, JHEP 04 (2022) 091.
\newblock \href {http://arxiv.org/abs/2111.06296} {\path{arXiv:2111.06296}},
  \href {https://doi.org/10.1007/JHEP04(2022)091}
  {\path{doi:10.1007/JHEP04(2022)091}}.

\bibitem{Baer:2013jla}
H.~Baer, V.~Barger, M.~Padeffke-Kirkland, X.~Tata, {Naturalness implies
  intra-generational degeneracy for decoupled squarks and sleptons}, Phys. Rev.
  D 89~(3) (2014) 037701.
\newblock \href {http://arxiv.org/abs/1311.4587} {\path{arXiv:1311.4587}},
  \href {https://doi.org/10.1103/PhysRevD.89.037701}
  {\path{doi:10.1103/PhysRevD.89.037701}}.

\bibitem{Baer:2019zfl}
H.~Baer, V.~Barger, D.~Sengupta, {Landscape solution to the SUSY flavor and CP
  problems}, Phys. Rev. Res. 1~(3) (2019) 033179.
\newblock \href {http://arxiv.org/abs/1910.00090} {\path{arXiv:1910.00090}},
  \href {https://doi.org/10.1103/PhysRevResearch.1.033179}
  {\path{doi:10.1103/PhysRevResearch.1.033179}}.

\bibitem{Baer:2012ts}
H.~Baer, V.~Barger, A.~Lessa, W.~Sreethawong, X.~Tata, {Wh plus missing-$E_T$
  signature from gaugino pair production at the LHC}, Phys. Rev. D 85 (2012)
  055022.
\newblock \href {http://arxiv.org/abs/1201.2949} {\path{arXiv:1201.2949}},
  \href {https://doi.org/10.1103/PhysRevD.85.055022}
  {\path{doi:10.1103/PhysRevD.85.055022}}.

\bibitem{ATLAS:2021moa}
G.~Aad, et~al., {Search for chargino\textendash{}neutralino pair production in
  final states with three leptons and missing transverse momentum in $\sqrt{s}
  = 13$~TeV pp collisions with the ATLAS detector}, Eur. Phys. J. C 81~(12)
  (2021) 1118.
\newblock \href {http://arxiv.org/abs/2106.01676} {\path{arXiv:2106.01676}},
  \href {https://doi.org/10.1140/epjc/s10052-021-09749-7}
  {\path{doi:10.1140/epjc/s10052-021-09749-7}}.

\bibitem{ATLAS:2018qmw}
M.~Aaboud, et~al., {Search for chargino and neutralino production in final
  states with a Higgs boson and missing transverse momentum at $\sqrt{s} = 13$
  TeV with the ATLAS detector}, Phys. Rev. D 100~(1) (2019) 012006.
\newblock \href {http://arxiv.org/abs/1812.09432} {\path{arXiv:1812.09432}},
  \href {https://doi.org/10.1103/PhysRevD.100.012006}
  {\path{doi:10.1103/PhysRevD.100.012006}}.

\bibitem{Carpenter:2023agq}
L.~M. Carpenter, H.~Gilmer, J.~Kawamura, T.~Murphy, {Taking aim at the
  wino-higgsino plane with the LHC} (9 2023).
\newblock \href {http://arxiv.org/abs/2309.07213} {\path{arXiv:2309.07213}}.

\bibitem{Baer:2018hpb}
H.~Baer, V.~Barger, J.~S. Gainer, D.~Sengupta, H.~Serce, X.~Tata, {LHC
  luminosity and energy upgrades confront natural supersymmetry models}, Phys.
  Rev. D 98~(7) (2018) 075010.
\newblock \href {http://arxiv.org/abs/1808.04844} {\path{arXiv:1808.04844}},
  \href {https://doi.org/10.1103/PhysRevD.98.075010}
  {\path{doi:10.1103/PhysRevD.98.075010}}.

\bibitem{Baer:2013xua}
H.~Baer, V.~Barger, P.~Huang, D.~Mickelson, A.~Mustafayev, W.~Sreethawong,
  X.~Tata, {Radiatively-driven natural supersymmetry at the LHC}, JHEP 12
  (2013) 013, [Erratum: JHEP 06, 053 (2015)].
\newblock \href {http://arxiv.org/abs/1310.4858} {\path{arXiv:1310.4858}},
  \href {https://doi.org/10.1007/JHEP12(2013)013}
  {\path{doi:10.1007/JHEP12(2013)013}}.

\bibitem{Matalliotakis:1994ft}
D.~Matalliotakis, H.~P. Nilles, {Implications of nonuniversality of soft terms
  in supersymmetric grand unified theories}, Nucl. Phys. B 435 (1995) 115--128.
\newblock \href {http://arxiv.org/abs/hep-ph/9407251}
  {\path{arXiv:hep-ph/9407251}}, \href
  {https://doi.org/10.1016/0550-3213(94)00487-Y}
  {\path{doi:10.1016/0550-3213(94)00487-Y}}.

\bibitem{Ellis:2002wv}
J.~R. Ellis, K.~A. Olive, Y.~Santoso, {The MSSM parameter space with
  nonuniversal Higgs masses}, Phys. Lett. B 539 (2002) 107--118.
\newblock \href {http://arxiv.org/abs/hep-ph/0204192}
  {\path{arXiv:hep-ph/0204192}}, \href
  {https://doi.org/10.1016/S0370-2693(02)02071-3}
  {\path{doi:10.1016/S0370-2693(02)02071-3}}.

\bibitem{Nath:1997qm}
P.~Nath, R.~L. Arnowitt, {Nonuniversal soft SUSY breaking and dark matter},
  Phys. Rev. D 56 (1997) 2820--2832.
\newblock \href {http://arxiv.org/abs/hep-ph/9701301}
  {\path{arXiv:hep-ph/9701301}}, \href
  {https://doi.org/10.1103/PhysRevD.56.2820}
  {\path{doi:10.1103/PhysRevD.56.2820}}.

\bibitem{Paige:2003mg}
F.~E. Paige, S.~D. Protopopescu, H.~Baer, X.~Tata, {ISAJET 7.69: A Monte Carlo
  event generator for pp, anti-p p, and e+e- reactions} (12 2003).
\newblock \href {http://arxiv.org/abs/hep-ph/0312045}
  {\path{arXiv:hep-ph/0312045}}.

\bibitem{Baer:2022qqr}
H.~Baer, V.~Barger, X.~Tata, K.~Zhang, {Prospects for Heavy Neutral SUSY HIGGS
  Scalars in the hMSSM and Natural SUSY at LHC Upgrades}, Symmetry 14~(10)
  (2022) 2061.
\newblock \href {http://arxiv.org/abs/2209.00063} {\path{arXiv:2209.00063}},
  \href {https://doi.org/10.3390/sym14102061} {\path{doi:10.3390/sym14102061}}.

\bibitem{Beenakker:1996ed}
W.~Beenakker, R.~Hopker, M.~Spira, {PROSPINO: A Program for the production of
  supersymmetric particles in next-to-leading order QCD} (11 1996).
\newblock \href {http://arxiv.org/abs/hep-ph/9611232}
  {\path{arXiv:hep-ph/9611232}}.

\bibitem{Baer:2017gzf}
H.~Baer, V.~Barger, J.~S. Gainer, M.~Savoy, D.~Sengupta, X.~Tata, {Aspects of
  the same-sign diboson signature from wino pair production with light
  higgsinos at the high luminosity LHC}, Phys. Rev. D 97~(3) (2018) 035012.
\newblock \href {http://arxiv.org/abs/1710.09103} {\path{arXiv:1710.09103}},
  \href {https://doi.org/10.1103/PhysRevD.97.035012}
  {\path{doi:10.1103/PhysRevD.97.035012}}.

\bibitem{Skands:2003cj}
P.~Z. Skands, et~al., {SUSY Les Houches accord: Interfacing SUSY spectrum
  calculators, decay packages, and event generators}, JHEP 07 (2004) 036.
\newblock \href {http://arxiv.org/abs/hep-ph/0311123}
  {\path{arXiv:hep-ph/0311123}}, \href
  {https://doi.org/10.1088/1126-6708/2004/07/036}
  {\path{doi:10.1088/1126-6708/2004/07/036}}.

\bibitem{Sjostrand:2006za}
T.~Sjostrand, S.~Mrenna, P.~Z. Skands, {PYTHIA 6.4 Physics and Manual}, JHEP 05
  (2006) 026.
\newblock \href {http://arxiv.org/abs/hep-ph/0603175}
  {\path{arXiv:hep-ph/0603175}}, \href
  {https://doi.org/10.1088/1126-6708/2006/05/026}
  {\path{doi:10.1088/1126-6708/2006/05/026}}.

\bibitem{Alwall:2011uj}
J.~Alwall, M.~Herquet, F.~Maltoni, O.~Mattelaer, T.~Stelzer, {MadGraph 5 :
  Going Beyond}, JHEP 06 (2011) 128.
\newblock \href {http://arxiv.org/abs/1106.0522} {\path{arXiv:1106.0522}},
  \href {https://doi.org/10.1007/JHEP06(2011)128}
  {\path{doi:10.1007/JHEP06(2011)128}}.

\bibitem{LHCHiggsCrossSectionWorkingGroup:2016ypw}
D.~de~Florian, et~al., {Handbook of LHC Higgs Cross Sections: 4. Deciphering
  the Nature of the Higgs Sector} 2/2017 (10 2016).
\newblock \href {http://arxiv.org/abs/1610.07922} {\path{arXiv:1610.07922}},
  \href {https://doi.org/10.23731/CYRM-2017-002}
  {\path{doi:10.23731/CYRM-2017-002}}.

\bibitem{Campbell:2011bn}
J.~M. Campbell, R.~K. Ellis, C.~Williams, {Vector boson pair production at the
  LHC}, JHEP 07 (2011) 018.
\newblock \href {http://arxiv.org/abs/1105.0020} {\path{arXiv:1105.0020}},
  \href {https://doi.org/10.1007/JHEP07(2011)018}
  {\path{doi:10.1007/JHEP07(2011)018}}.

\bibitem{deFavereau:2013fsa}
J.~de~Favereau, C.~Delaere, P.~Demin, A.~Giammanco, V.~Lema\^\i{}tre,
  A.~Mertens, M.~Selvaggi, {DELPHES 3, A modular framework for fast simulation
  of a generic collider experiment}, JHEP 02 (2014) 057.
\newblock \href {http://arxiv.org/abs/1307.6346} {\path{arXiv:1307.6346}},
  \href {https://doi.org/10.1007/JHEP02(2014)057}
  {\path{doi:10.1007/JHEP02(2014)057}}.

\bibitem{Krohn_jet_trim}
D.~Krohn, J.~Thaler, L.-T. Wang, {Jet Trimming}, JHEP 02 (2010) 084.
\newblock \href {http://arxiv.org/abs/0912.1342} {\path{arXiv:0912.1342}},
  \href {https://doi.org/10.1007/JHEP02(2010)084}
  {\path{doi:10.1007/JHEP02(2010)084}}.

\bibitem{Baer:2013yha}
H.~Baer, V.~Barger, P.~Huang, D.~Mickelson, A.~Mustafayev, W.~Sreethawong,
  X.~Tata, {Same sign diboson signature from supersymmetry models with light
  higgsinos at the LHC}, Phys. Rev. Lett. 110~(15) (2013) 151801.
\newblock \href {http://arxiv.org/abs/1302.5816} {\path{arXiv:1302.5816}},
  \href {https://doi.org/10.1103/PhysRevLett.110.151801}
  {\path{doi:10.1103/PhysRevLett.110.151801}}.

\bibitem{Barr:2003rg}
A.~Barr, C.~Lester, P.~Stephens, {m(T2): The Truth behind the glamour}, J.
  Phys. G 29 (2003) 2343--2363.
\newblock \href {http://arxiv.org/abs/hep-ph/0304226}
  {\path{arXiv:hep-ph/0304226}}, \href
  {https://doi.org/10.1088/0954-3899/29/10/304}
  {\path{doi:10.1088/0954-3899/29/10/304}}.

\bibitem{Barger:1983jx}
V.~D. Barger, A.~D. Martin, R.~J.~N. Phillips, {Evidence for the t Quark in
  anti-p p Collider Data}, Phys. Lett. B 125 (1983) 339.
\newblock \href {https://doi.org/10.1016/0370-2693(83)91297-2}
  {\path{doi:10.1016/0370-2693(83)91297-2}}.

\bibitem{Barnett:1987kn}
R.~M. Barnett, J.~F. Gunion, H.~E. Haber, {Gluino Decay Patterns and
  Signatures}, Phys. Rev. D 37 (1988) 1892.
\newblock \href {https://doi.org/10.1103/PhysRevD.37.1892}
  {\path{doi:10.1103/PhysRevD.37.1892}}.

\bibitem{Baer:1989hr}
H.~Baer, X.~Tata, J.~Woodside, {Gluino Cascade Decay Signatures at the Tevatron
  Collider}, Phys. Rev. D 41 (1990) 906--915.
\newblock \href {https://doi.org/10.1103/PhysRevD.41.906}
  {\path{doi:10.1103/PhysRevD.41.906}}.

\bibitem{Baer:1991xs}
H.~Baer, X.~Tata, J.~Woodside, {Multi - lepton signals from supersymmetry at
  hadron super colliders}, Phys. Rev. D 45 (1992) 142--160.
\newblock \href {https://doi.org/10.1103/PhysRevD.45.142}
  {\path{doi:10.1103/PhysRevD.45.142}}.

\bibitem{Barnett:1993ea}
R.~M. Barnett, J.~F. Gunion, H.~E. Haber, {Discovering supersymmetry with like
  sign dileptons}, Phys. Lett. B 315 (1993) 349--354.
\newblock \href {http://arxiv.org/abs/hep-ph/9306204}
  {\path{arXiv:hep-ph/9306204}}, \href
  {https://doi.org/10.1016/0370-2693(93)91623-U}
  {\path{doi:10.1016/0370-2693(93)91623-U}}.

\bibitem{Baer:1995va}
H.~Baer, C.-h. Chen, F.~Paige, X.~Tata, {Signals for minimal supergravity at
  the CERN large hadron collider. 2: Multi - lepton channels}, Phys. Rev. D 53
  (1996) 6241--6264.
\newblock \href {http://arxiv.org/abs/hep-ph/9512383}
  {\path{arXiv:hep-ph/9512383}}, \href
  {https://doi.org/10.1103/PhysRevD.53.6241}
  {\path{doi:10.1103/PhysRevD.53.6241}}.

\bibitem{Read_2002}
A.~L. Read, Presentation of search results: the ${CL_s}$ technique, J. Phys. G
  28 (2002) 2693.
\newblock \href {https://doi.org/10.1088/0954-3899/28/10/313}
  {\path{doi:10.1088/0954-3899/28/10/313}}.

\bibitem{Cowan_2011}
G.~Cowan, K.~Cranmer, E.~Gross, O.~Vitells, Asymptotic formulae for
  likelihood-based tests of new physics, Eur.Phys. J. C 71 (2011).
\newblock \href {http://arxiv.org/abs/1007.1727} {\path{arXiv:1007.1727}},
  \href {https://doi.org/10.1140/epjc/s10052-011-1554-0}
  {\path{doi:10.1140/epjc/s10052-011-1554-0}}.

\bibitem{Baer:2006rs}
H.~Baer, X.~Tata, {Weak scale supersymmetry: From superfields to scattering
  events}, Cambridge University Press, 2006.

\bibitem{Baer:1990in}
H.~Baer, X.~Tata, J.~Woodside, {${Z}^{0}+\mathrm{jets}+{p}_{T}$ events as a
  signal for supersymmetry at the Fermilab Tevatron collider}, Phys. Rev. D 42
  (1990) 1450--1454.
\newblock \href {https://doi.org/10.1103/PhysRevD.42.1450}
  {\path{doi:10.1103/PhysRevD.42.1450}}.

\bibitem{Baer:2014yta}
H.~Baer, V.~Barger, D.~Mickelson, A.~Mustafayev, X.~Tata, {Physics at a
  Higgsino Factory}, JHEP 06 (2014) 172.
\newblock \href {http://arxiv.org/abs/1404.7510} {\path{arXiv:1404.7510}},
  \href {https://doi.org/10.1007/JHEP06(2014)172}
  {\path{doi:10.1007/JHEP06(2014)172}}.

\bibitem{Moortgat-Pick:2015lbx}
A.~Arbey, et~al., {Physics at the e+ e- Linear Collider}, Eur. Phys. J. C
  75~(8) (2015) 371.
\newblock \href {http://arxiv.org/abs/1504.01726} {\path{arXiv:1504.01726}},
  \href {https://doi.org/10.1140/epjc/s10052-015-3511-9}
  {\path{doi:10.1140/epjc/s10052-015-3511-9}}.

\bibitem{Baer:2016new}
H.~Baer, M.~Berggren, K.~Fujii, S.-L. Lehtinen, J.~List, T.~Tanabe, J.~Yan,
  {Naturalness and light higgsinos: A powerful reason to build the ILC}, PoS
  ICHEP2016 (2016) 156.
\newblock \href {http://arxiv.org/abs/1611.02846} {\path{arXiv:1611.02846}},
  \href {https://doi.org/10.22323/1.282.0156} {\path{doi:10.22323/1.282.0156}}.

\bibitem{Lehtinen:2017vdt}
S.-L. Lehtinen, H.~Baer, M.~Berggren, J.~List, K.~Fujii, J.~Yan, T.~Tanabe,
  {Naturalness and light Higgsinos: why ILC is the right machine for SUSY
  discovery}, PoS EPS-HEP2017 (2017) 306.
\newblock \href {http://arxiv.org/abs/1710.02406} {\path{arXiv:1710.02406}},
  \href {https://doi.org/10.22323/1.314.0306} {\path{doi:10.22323/1.314.0306}}.

\bibitem{Baer:2019gvu}
H.~Baer, M.~Berggren, K.~Fujii, J.~List, S.-L. Lehtinen, T.~Tanabe, J.~Yan,
  {ILC as a natural SUSY discovery machine and precision microscope: From light
  Higgsinos to tests of unification}, Phys. Rev. D 101~(9) (2020) 095026.
\newblock \href {http://arxiv.org/abs/1912.06643} {\path{arXiv:1912.06643}},
  \href {https://doi.org/10.1103/PhysRevD.101.095026}
  {\path{doi:10.1103/PhysRevD.101.095026}}.

\bibitem{Baer:2020sgm}
H.~Baer, V.~Barger, S.~Salam, D.~Sengupta, X.~Tata, {The LHC higgsino discovery
  plane for present and future SUSY searches}, Phys. Lett. B 810 (2020) 135777.
\newblock \href {http://arxiv.org/abs/2007.09252} {\path{arXiv:2007.09252}},
  \href {https://doi.org/10.1016/j.physletb.2020.135777}
  {\path{doi:10.1016/j.physletb.2020.135777}}.

\end{thebibliography}
\bibliographystyle{elsarticle-num}

\end{document}